\documentclass[lettersize,journal]{IEEEtran}
\usepackage{amsmath,amsfonts,amssymb}
\usepackage{algorithm}
\usepackage{array}
\usepackage[caption=false,font=normalsize,labelfont=sf,textfont=sf]{subfig}
\usepackage{textcomp}
\usepackage{stfloats}
\usepackage{float}
\usepackage{url}
\usepackage{verbatim}
\usepackage{cite}
\usepackage{algorithmicx}
\usepackage{algpseudocode}
\usepackage{graphicx,color}
\newtheorem{assumption}{Assumption}
\begin{document}

\title{AdRo-FL: Informed and Secure Client Selection for Federated Learning in the Presence of Adversarial Aggregator}

\author{Md. Kamrul Hossain, Walid Aljoby, Anis Elgabli, Ahmed M. Abdelmoniem, Khaled A. Harras
\thanks{Md. Kamrul Hossain, Walid Aljoby and Anis Elgabli are with College of Computing and Mathematics, King Fahd University of Petroleum \& Minerals, Dhahran, Saudi Arabia (email: g202215400@kfupm.edu.sa; waleed.gobi@kfupm.edu.sa; anis.elgabli@kfupm.edu.sa).}
\thanks{Ahmed M. Abdelmoniem is with School of Electronic Engineering and Computer Science, Queen Mary University of London, United Kingdom (email: ahmed.sayed@qmul.ac.uk).}
\thanks{Khaled A. Harras is with Department of Computer Science, Carnegie Mellon University, United States (email: kharras@cs.cmu.edu).}
\thanks{Corresponding Author: Walid Aljoby.}

}

\markboth{Journal of \LaTeX\ Class Files,~Vol.~14, No.~8, August~2021}%
{Shell \MakeLowercase{\textit{et al.}}: A Sample Article Using IEEEtran.cls for IEEE Journals}


\maketitle

\begin{abstract}
%Federated Learning (FL) is collaborative training of machine learning models without exposing local client data. Even though clients only share model updates with the aggregator, studies have shown that an aggregator can infer information about clients local data from the models. Secure Aggregation (SA) is often used to prevent this, which ensures that individual client updates remain private while they’re being sent to the central aggregator. But recent research has uncovered a significant vulnerability that showed that the aggregator can manipulate client selection to infer private model updates even when SA is used, enabling what’s known as a biased selection attack (BSA). Although verifiable random selection of clients can prevent this attack, it does not allow informed client selection. 
%In this paper, we present Adversarial Robust Federated Learning (AdRo-FL) that performs informed client selection while protecting against biased selection attack by an adversarial aggregator. 

Federated Learning (FL) enables collaborative model training without exposing clients' local data. While clients only share model updates with the aggregator, studies reveal that malicious aggregators can infer sensitive information from these updates. Secure Aggregation (SA) protects individual updates during transmission; however, recent work demonstrates a critical vulnerability where adversarial aggregators manipulate client selection to bypass SA protections, constituting a Biased Selection Attack (BSA). Although verifiable random selection prevents BSA, it precludes informed client selection essential for FL performance.

We propose Adversarial Robust Federated Learning (AdRo-FL), which simultaneously enables: (1) informed client selection based on client utility, and (2) robust defense against BSA maintaining privacy-preserving aggregation. 
%Our approach maintains privacy-preserving aggregation while ensuring selection integrity against adversarial manipulation.
%AdRo-FL implements two client selection frameworks for two different client settings in FL. The first framework assumes that clients belong to clusters based on mutual trust such as different branches of an organization. The second framework assumes that clients are distributed and no trust exists between the clients.  
AdRo-FL implements two client selection frameworks tailored for distinct settings in FL. The first framework assumes clients are grouped into clusters based on mutual trust, such as different branches of an organization. The second framework handles distributed clients where no trust relationships exist between them.
For the cluster-oriented setting, we propose a novel defense against BSA by (1) enforcing a minimum client selection quota from each cluster, supervised by a cluster-head in every round, and (2) introducing a client utility function to prioritize efficient clients. For the distributed setting, we design a two-phase selection protocol: first, the aggregator selects the top 80\% of clients based on our utility-driven ranking; then, a verifiable random function (VRF) ensures transparent, auditable, and BSA-resistant final selection from this pool. 
%For the cluster-oriented setting, we strengthen defenses against BSA by enforcing a minimum number of selected clients from each cluster supervised by a cluster-head in every training round. We also perform informed client selection using client utility. As for the distributed, non-cluster-oriented setting, we allow the aggregator to perform an initial selection by selecting the top 80\% clients based on the utility value. Then we apply verifiable random function (VRF) on the primary selection pool to select desired number of clients for participating in the FL. VRF makes the selection transparent and verifiable and BSA-proof. 
AdRo-FL also applies quantization to reduce communication overhead and sets strict transmission deadlines to improve energy efficiency. AdRo-FL achieves up to $1.85\times$ faster convergence (time-to-accuracy) and up to $1.06\times$ higher final accuracy compared to insecure baselines across multiple datasets. 
%We provide our implementation codes at GitHub: https://github.com/Muhammadkamrul/SecureInformedFL
\end{abstract}


\maketitle

\section{INTRODUCTION}
\label{intro}
\IEEEPARstart{F}{ederated} Learning (FL)~\cite{zhang2021survey} has gained significant attention as a decentralized paradigm for collaborative machine learning, where multiple clients train a shared model without exchanging raw data. This approach is especially suitable for privacy-sensitive applications such as healthcare, finance, and industrial IoT (IIoT), where data confidentiality is paramount. In FL, clients perform local training on their private data and periodically send model updates to a central aggregator, which aggregates them to refine the global model. While FL enhances privacy by design, it remains susceptible to adversarial threats, particularly when the central aggregator behaves maliciously.

One of the critical vulnerabilities of FL is its exposure to aggregators, which can exploit model updates to extract sensitive client information~\cite{le2023privacy, CAI2024102420, 9464278}. Several privacy-preserving techniques have been proposed to mitigate this risk, including Differential Privacy (DP)~\cite{fi15090310,RODRIGUEZBARROSO2020270} and Secure Aggregation (SA)~\cite{10.1145/3133956.3133982}. DP introduces noise into model updates before transmission, providing strong privacy guarantees. However, this noise reduces model accuracy, making it less suitable for scenarios that require high-performance learning. SA, on the other hand, enables clients to encrypt their updates before transmission, ensuring that only aggregated results are revealed to the aggregator. This makes SA a more practical choice in many FL settings, particularly when privacy must be preserved without significantly compromising model performance.

Despite its advantages, SA itself is not entirely secure. Recent studies have shown that biased selection attacks (BSA)~\cite{nguyen2022blockchain,298104} can undermine SA by allowing a adversarial aggregator to manipulate client selection, thereby inferring individual model updates. 
BSA can manifest in two forms: \textit{non-colluding} attacks, where an adversarial aggregator isolates a specific client to obtain its model update, and \textit{colluding} attacks, where an adversarial aggregator colludes with a subset of clients to uncover the model update of the target client.
%BSA can manifest in two forms: \textbf{non-colluding} attacks, where an honest-but-curious (who follows the protocol) aggregator isolates a specific client to analyze its updates, and \textbf{colluding attacks}, where an adversarial (who does not follow the protocol) aggregator conspires with a subset of clients in order to uncover sensitive information about the target client. 
This vulnerability highlights the need for a resilient FL framework that prevents the exploitation of SA while maintaining computational efficiency as well as model quality.

To defend against BSA, researchers have recently proposed using verifiable random functions (VRF)~\cite{298104, nguyen2022blockchain} to allow each client to independently verify whether the selection process was genuinely random. A technical overview of VRFs is provided in the appendix~\ref{appendix:vrf}.
%(a technical overview of VRFs is provided in the appendix~\ref{appendix:vrf}). 
While truly random client selection can effectively disrupt BSA, this approach lacks the ability to make informed decisions when selecting clients. Informed client selection~\cite{li2024comprehensive} remains crucial for achieving faster convergence and improving energy efficiency in FL.

%Although truly random client selection can effectively disrupt any BSA, this approach, however, falls short in making informed decisions when choosing clients.
%Informed client selection~\cite{li2024comprehensive} is crucial for faster convergence and energy efficiency in FL. 

In this paper, to address this problem, we propose Adversarial Robust Federated Learning \textbf{(AdRo-FL)} that allows informed client selection while protecting against biased selection attack by an adversarial aggregator. 
The development of AdRo-FL is motivated by several key challenges in FL. First, while SA prevents direct leakage of model updates, it remains vulnerable to BSA that allows adversarial aggregator to infer local model update. Second, existing defenses that rely on random client selection often degrade performance by disregarding the quality of client contributions. There is a notable lack of solutions that effectively balance resilience to BSA with optimized client selection based on data utility and system efficiency. Lastly, many real-world deployments, such as smart cities, industrial IoT, and finance, exhibit natural client clustering based on organizational and administrative trust, which remains underutilized in current approaches.

\begin{figure*}[hbt!]
    \includegraphics[width=.9\linewidth]{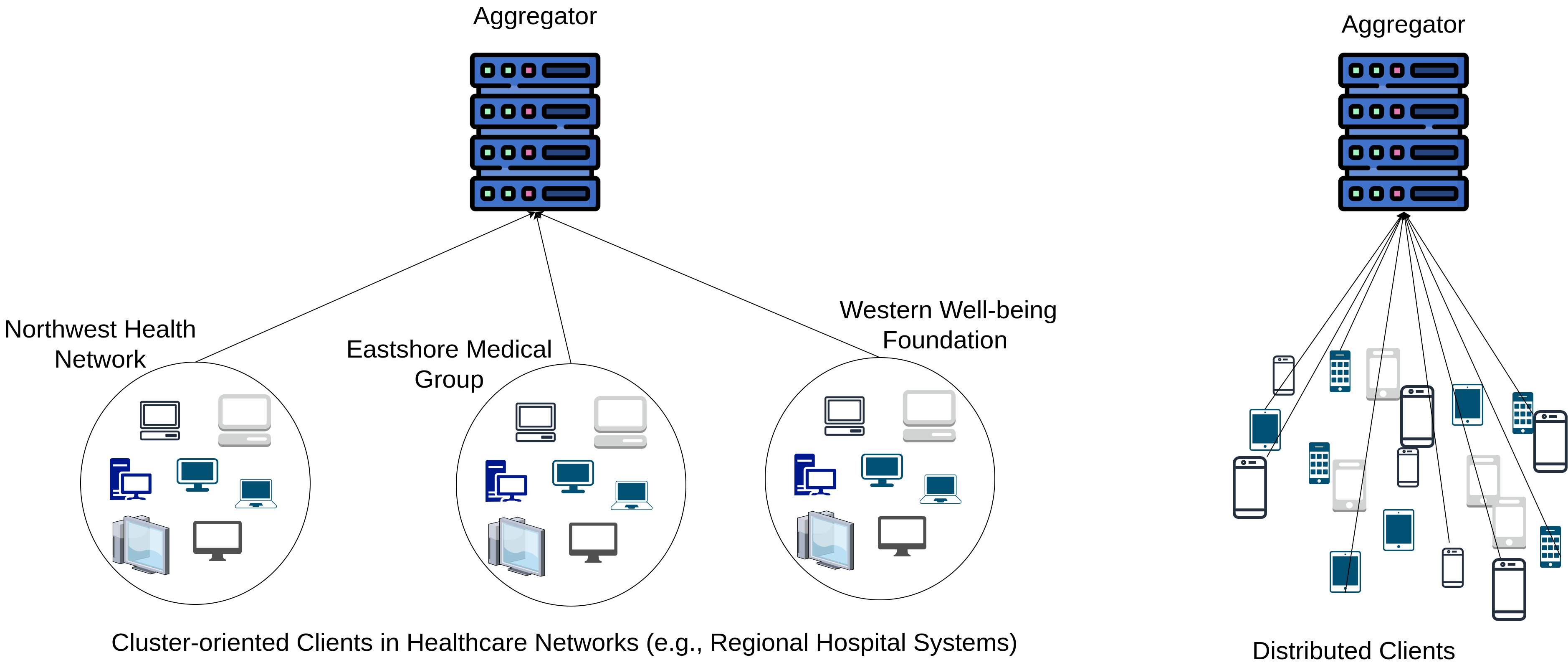}
    \caption{Federated Learning client settings.}
    \label{fl_env}
\end{figure*}

AdRo-FL considers two different client settings commonly found in real-world environments. The first setting assumes that clients belong to clusters where there exist mutual trusts within, such as different branches of an organization (Fig. \ref{fl_env}, Healthcare Networks). The second setting assumes that the clients are distributed, non-cluster-oriented and that there is no trust between the clients (Fig. \ref{fl_env}, Cellphone Users). 

For the cluster-oriented setting, we assume a cluster-head per cluster that intermediates communications between the cluster and the aggregator. We enforce that either a minimum number \(C\) of clients must be selected from a cluster or no clients will join from that cluster. That is, either the aggregator will select \(\geq C\) clients from a cluster or none. The cluster-head will supervise and verify this process in every training round. This process guarantees protection against BSA and is detailed in section \ref{prevent_bsa_in_cluster}.

%This cluster-oriented client setting is inspired by real-world IoT ecosystems, where organizations often have structured trust relationships. For instance, in smart cities, municipal departments collaborate within their administrative boundaries while engaging with external agencies for broader urban optimization. Similarly, in industrial IoT (IIoT), factories within the same corporate group trust each other but interact cautiously with external suppliers. Likewise, in financial networks, banks within the same institution may collaborate via FL to detect fraudulent transactions while safeguarding their proprietary data. These real-world examples underscore the relevance of cluster-oriented FL in ensuring privacy while enabling collaborative learning.

As for the distributed, non-cluster-oriented setting, the client selection is completed in two phases. In the first phase, clients compute a utility value and sign it using their private keys. These signed values are sent to the aggregator, which ranks them, compresses the sorted list and signatures, and broadcasts both. Clients in the top 80\% proceed to the next selection phase as detailed in section \ref{informed_selection_VRF}. This two level client selection strategy ensures informed as well as secure client selection. The initial selection retains best clients based on their utility and the final selection ensures verifiable random selection to prevent BSA.

Beyond privacy protection, AdRo-FL enhances FL's efficiency through combining several key approaches:
\begin{itemize}
  \item \textbf{Utility-based Client Selection}: It selects clients based on a combination of local loss, gradient norm and sample size, prioritizing clients that contribute the most to global model.
  \item \textbf{Deadline-aware Selection}: To ensure timely updates and energy efficiency in resource-constrained IoT environments, it incorporates a deadline-based transmission mechanism, allowing clients to only participate within a specified time constraints.
  \item \textbf{Quantization for Communication Efficiency}: To reduce the communication overhead, it employs quantization, compressing model updates before transmission.
\end{itemize}

To evaluate the effectiveness of AdRo-FL, we conduct extensive experiments on four benchmark datasets: MNIST, FMNIST, SVHN, and CIFAR-10. We simulate non-IID data distribution using Dirichlet distribution\cite{jimenez2024fedartml}, closely mirroring real-world FL scenarios. Our results demonstrate that AdRo-FL significantly improves training accuracy and energy efficiency compared to insecure client selection methods such as Oort~\cite{273723} which is a state-of-the-art client selection method in FL.

Below we summarize our contributions:
\begin{itemize}
  \item We propose AdRo-FL that supports both cluster-oriented and non-clustered client settings. In cluster-based environments, AdRo-FL leverages intra-cluster trust to defend against BSA while enabling informed client selection. For non-clustered, distributed setting, it introduces a two-phase client filtering mechanism to ensure secure and informed client selection.
  \item We propose a client utility measurement function combining gradient norm, local loss, and sample size. Additionally, clients are filtered based on a transmission deadline to improve energy efficiency. We apply 8-bit quantization to reduce communication overhead and enhance scalability in bandwidth-constrained IoT systems without affecting performance.
  \item We conduct comprehensive experiments on MNIST, FMNIST, SVHN, and CIFAR10 datasets by using Dirichlet distribution with high heterogeneity, demonstrating superior performance over insecure client selection baselines.
\end{itemize}

%The rest of this paper is organized as follows. Section~\ref{threatmodel} illustrates the threat model. Section~\ref{RW} reviews related work. Section~\ref{Methodology_overview} gives an overview of the proposed AdRo-FL. Section~\ref{Methodology_AdRo-FL_defense} presents AdRo-FL's defense mechanism against BSA. Section \ref{Methodology_Efficient_Client_Selection} presents how AdRo-FL selects clients to optimize convergence speed and energy consumption. Section \ref{ER} describes the experimental results. Finally, Section~\ref{C} concludes the article.

\section{THREAT MODEL}
\label{threatmodel}

\begin{figure}[htb!]
\centering
\includegraphics[width=\columnwidth]{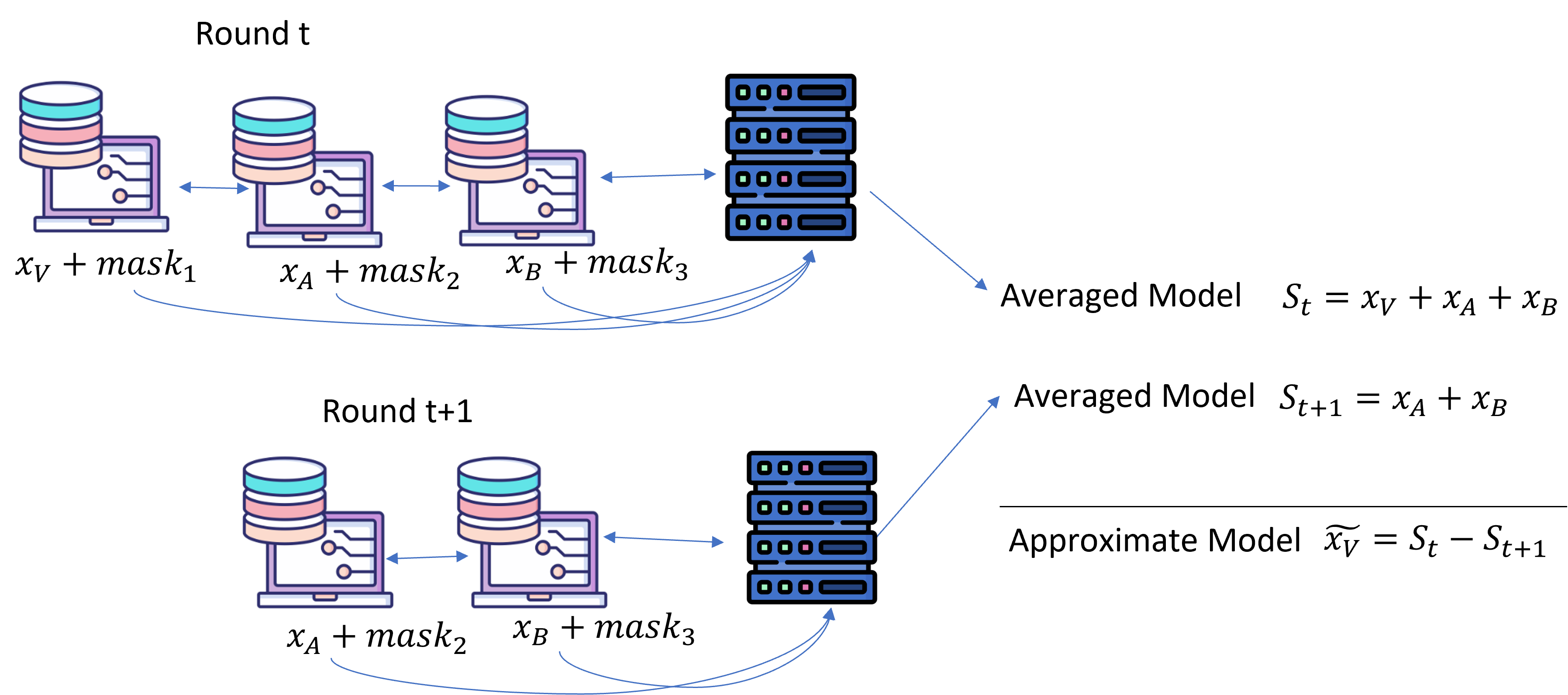}
\caption{Non-colluding biased selection attack.}
\label{bsa1}
\end{figure}

\begin{figure}[htb!]
\centering
\includegraphics[width=\columnwidth]{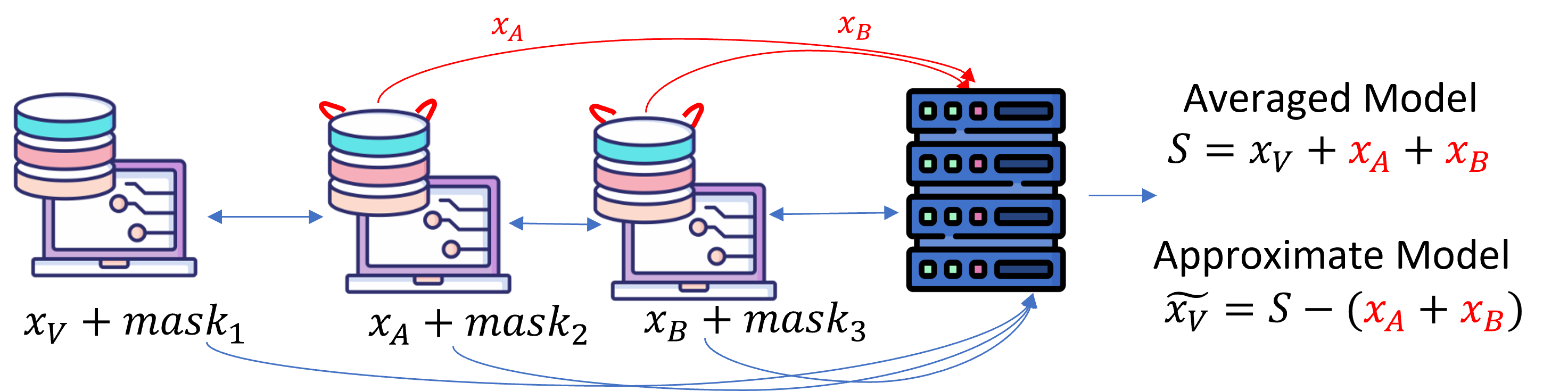}
\caption{Colluding biased selection attack.}
\label{bsa2}
\end{figure}

This section describes how a BSA is executed in SA. In \cite{nguyen2022blockchain}, two types of attacks are described. The first, a non-colluding attack (Fig. \ref{bsa1}), assumes no collusion between clients and the aggregator, nor the creation of fake clients by the aggregator. Here, the aggregator uses statistical inference by subtracting consecutive aggregated updates. Suppose, in round \(t\), the aggregator selects the victim client \(V\) and a subset of other clients ($A$ and $B$), resulting in an aggregated update \(S_t = x_V + x_A + x_B\). Here $x$ represents the local model update. In the following round \(t+1\), the aggregator reselect the same clients, excluding \(V\), obtaining \(S_{t+1} = x_A + x_B\). By estimating \(S_{t+1} - S_t\), the aggregator approximates \(V\)'s model \(x_V\). However, the work ~\cite{nguyen2022blockchain} mentioned some preconditions to make this attack feasible, such as the aggregator should either send same global model to the selected clients at the rounds \(t\) and \(t+1\) or do this attack at the late stage when the global model has already converged. It was also assumed that the clients (other than the victim) do not change the algorithm and local data between the rounds \(t\) and \(t+1\). 

The second type, a colluding attack (Fig. \ref{bsa2}), relaxes these preconditions by assuming that some clients collude with the aggregator by sharing their model updates. Here, the aggregator aggregates the colluding clients’ updates ($x_A, x_B$) alongside victim \(V\)'s, obtaining \(S = x_V + x_A + x_B\). Since the aggregator knows \(x_A\) and \(x_B\) through colluding clients, it can approximate \(x_V\) using \(x_V \approxeq S - (x_A + x_B)\).

\section{RELATED WORK}
\label{RW}
\subsection{Insecure Informed Client Selection}
Client selection plays a crucial role in improving global model convergence and energy efficiency in FL. While random client selection is conventionally used, researchers have explored more intelligent informed selection strategies. There are two dimensions of this. A group of works\cite{li2024node, tang2021fedgp, wu2022node} explored ways to improve global model convergence time and accuracy without considering resource efficiency while another group of works explored improving both the convergence and training efficiency in terms of resource consumption~\cite{10534777, 9846900, 10589575, 10145997}.

Most of the works that employ client selection for improved global model convergence leverages metrics such as local loss, gradient norm and training duration. For instance,~\cite{marnissi2024client} prioritizes clients with the highest L2 norm of gradients, under the assumption that these clients contribute most to the global model's learning progress. However, relying solely on gradient magnitude may oversimplify client importance, neglecting critical aspects such as data diversity and quality, which are essential for robust FL. Similarly,~\cite{cho2022towards} proposes the Power-of-Choice, where clients with higher local losses are favored to accelerate convergence. While this approach can improve accuracy, it introduces the risk of model bias, as clients with extreme loss values may disproportionately influence learning. Oort~\cite{273723} also showed that selecting clients with higher training loss can accelerate convergence. FairDPFL-SCS~\cite{SABAH2025102756} selects different subset of clients in each FL round while prioritizing the selection of new clients. The work~\cite{GUO2024102549} proposed an auction-based client selection mechanism for online FL to select better clients while compensating clients for their participation.

In addition to client selection for faster convergence, energy efficiency remains a major concern in FL, especially in resource-constrained environments such as IoT networks and mobile edge computing. Several strategies have been proposed to optimize energy consumption, including message compression, accuracy relaxation, and client dropout mechanisms. In~\cite{10643330}, an early stopping mechanism is introduced to terminate training when the global model begins to oscillate among local optima, thereby reducing unnecessary communication rounds. Another approach in~\cite{DBLP:journals/corr/Dettmers15} utilizes 8-bit approximations instead of traditional 32-bit gradient representations, effectively reducing communication overhead without significantly compromising accuracy. Additionally, PyramidFL~\cite{li2022pyramidfl} employs a trade-off strategy that allows slightly lower accuracy in favor of better communication efficiency by training sub-optimal solutions during local model updates.

Yet another group of work leverages client data and resource homogeneity to cluster them together to facilitate improved client selection and model convergence. For example, the work \cite{briggs2020federated} clustered clients based on their local model weights. \cite{9610118} does client clustering based on location and data similarity. Another work called `Auxo'\cite{10.1145/3620678.3624651}, groups clients with statistically similar data distributions, called `cohorts'. Yet another work \cite{10083200} clustered clients by training duration. The clusters are subsequently scheduled across training rounds to minimize the total number of iterations needed to reach a desired accuracy level. 

While the above mentioned client selection methods are effective in terms of accuracy and energy efficiency, they are insecure against an honest-but-curious aggregator and an adversarial aggregator.

\subsection{Secure Uninformed Client Selection}
Various techniques have been developed to allow a aggregator to obtain the sum of clients model updates without access to individual model update. However, many rely on assumptions that limit their practicality, such as an honest-but-curious aggregator or a trusted third party~\cite{MLSYS2022_6c44dc73, 10.5555/3310435.3310586}. Secure Aggregation (SA) is one of the few available protocols that were designed to handle a potentially malicious aggregator. SA~\cite{10.1145/3133956.3133982} has been widely adopted as a countermeasure, enabling encrypted aggregation of client updates to obscure individual contributions. However, SA alone is insufficient, as recent research has demonstrated its susceptibility to biased selection attacks (BSA)~\cite{nguyen2022blockchain}. These attacks allow a malicious aggregator to manipulate client selection in order to infer sensitive model updates, posing a significant threat to data privacy. Additionally, there is an inherent problem with SA, it can handle only a limited number of clients colluding with the aggregator; beyond this threshold, the adversary could reconstruct a client’s update~\cite{10.1145/3133956.3133982}. To strengthen SA, distributed DP~\cite{10.1145/3627703.3629559} adds another layer of privacy to the computed sum by combining DP noise addition with SA, offering robust privacy guarantees. Nevertheless, both SA and distributed DP remain susceptible to BSA when a majority of participants are dishonest. To mitigate BSA,~\cite{nguyen2022blockchain} introduces a blockchain-based random client selection strategy, leveraging blockchain’s transparency to ensure unbiased client participation. While this approach enhances trustworthiness, it introduces scalability challenges due to blockchain’s inherent limitations in transaction speed and data throughput, making it impractical for large-scale FL deployments. Additionally, random client selection overlooks key factors such as client utility which makes informed client selection impossible.

\subsection{Secure and Informed Client Selection}
The Lotto~\cite{298104} approach uses VRF to randomly select clients from a client-pool. The clients in the client-pool were chosen based on some desired client selection criteria where a adversarial aggregator was involved. One of the limitation of this work is that the adversarial aggregator may attempt to manipulate the process of building the client-pool by excluding honest clients which will enable a BSA. Moreover, it fails when the majority of clients are colluding while only the minority are honest. Another limitation of this work is that when the aggregator does a primary selection, it goes without a verification from client side before applying verifiable randomness based final selection. The adversarial aggregator may manipulate this to do a BSA by sending forged list to the clients. Moreover, the VRF \(alpha\) was chosen to be a deterministic number which gives the adversarial aggregator more control to create Sybils\cite{fung2020limitations} in the selection pool. To the best of our knowledge, this remain the only work specifically addressing BSA in SA, underscoring the need for more research on this area. Our proposed AdRo-FL offers several advantages over Lotto. Firstly, we do not use deterministic \(alpha\) value for the VRF (as Lotto does) which makes it more difficult for the adversarial aggregator to do BSA. Secondly, we include sign verification for the fist level selection of clients which makes the adversarial aggregator unable to forge the utility based selection process. Lotto does not address this vulnerability. Lastly, Lotto did not consider cluster-oriented client setting as a separate problem. We identify the existence of cluster-oriented clients setting and take advantage of the inherent trust of clusters to prevent BSA by an adversarial aggregator.

To summarize, existing works tend to focus on either privacy or model convergence and energy efficiency, but very few address all these challenges simultaneously. In real-world FL deployments, these factors are deeply interconnected, particularly in highly heterogeneous and resource-limited environments.

\section{\textnormal{AdRo-FL} Overview}
\label{Methodology_overview}
AdRo-FL aims to defend against BSA attempted by an adversarial aggregator while allowing informed client selection for improved global model and faster convergence. It takes advantage of the client's local loss and weighted gradient norm to determine client's utility. In addition, clients should be able to transmit their payloads within a specific deadline which helps faster transmission as well as less energy consumption. In addition, AdRo-FL incorporates quantization to reduce the payload size, improving communication efficiency in bandwidth-constrained networks. AdRo-FL implements two different frameworks to suit different client setting of real world environment to defend against BSA.

\begin{enumerate}
\item Informed client selection for cluster-oriented clients.
\item Informed client selection for non-cluster-oriented, distributed clients.
\end{enumerate}

In Section \ref{Methodology_AdRo-FL_defense}, we discuss the details of the client privacy part, which is our defense approach against the BSA in the two settings mentioned above. 

In Section \ref{Methodology_Efficient_Client_Selection}, we discuss the details of the optimization part, which includes the client utility that allows informed client selection, deadline-based client selection as well as quantization of the client's payload.

\section{\textnormal{AdRo-FL}: Defense against the BSA}
\label{Methodology_AdRo-FL_defense}

\subsection{\textbf{Informed client selection for cluster-oriented clients}}
This section describes a real-world client setting in which clients are grouped under a unit that we call a cluster. There is trust within each cluster but not outside. That is, a client from one cluster does not trust a client from another cluster. One of the advantages we get in this setting is that we don't need to use randomness to prevent BSA from adversarial aggregator. We leverage the intra-cluster trust to prevent BSA. We describe this framework in detail in the following subsections.

\subsubsection{Rationale for Clustering}
In many real-world deployments of federated learning, such as within corporate ecosystems, municipal infrastructures, or financial institutions, clients often exhibit natural grouping based on ownership or administrative control. For example, Smart city applications might cluster clients under municipal departments (e.g. traffic, utilities, public safety), Industrial IoT settings could group devices deployed across different factory floors of the same company, and healthcare networks might cluster devices or data silos within the same hospital system. A real example is the federated learning initiative for breast density classification involving seven clinical institutions worldwide\cite{10.1007/978-3-030-60548-3_18}. Each institution (cluster) trained local models on their proprietary data and shared model updates with a central aggregator. This collaborative approach allowed the development of a robust global model without exchanging sensitive patient data, adhering to privacy regulations such as HIPAA and GDPR~\cite{said2023hipaa}. These groupings provide a natural trust boundary where clients within the same cluster are assumed to be under a common administrative domain.

\subsubsection{Trust Assumptions}
Although clients in the same cluster do not share raw data or model updates with each other, they trust that none of them will collude with the aggregator or at least there will be a minority trusted clients withing each cluster. This assumption is vital for defending against BSA under SA. It ensures that the privacy threshold mechanism—requiring at least $C$ clients from each participating cluster—functions effectively which we describe in section \ref{prevent_bsa_in_cluster}.

\subsubsection{Cluster Formation Mechanism}
Cluster formation in our setting is declarative and administrative rather than algorithmic, with clusters defined during deployment based on existing organizational structures. Each client is assigned a signed cluster identifier by a central certificate authority (CA) or a trusted cluster head. This identifier supports cluster membership verification and enforces a privacy threshold that ensures a minimum number of clients participate in each aggregation round. This aligns with real-world workflows where devices are centrally registered, authenticated, and managed.

%\subsubsection{Practicality and Deployment Implications}
%This cluster-oriented model is particularly suitable for cross-silo federated learning settings where each cluster represents a silo such as a company branch, a department, or a local administrative region. In contrast to cross-device FL, where trust assumptions are weaker and devices are more heterogeneous, cross-silo environments naturally support the trust requirements. Moreover, by avoiding intra-cluster data sharing and requiring only metadata (such as loss values or gradient norms) to be reported to the aggregator via the cluster head, AdRo-FL ensures that privacy boundaries are respected even within a trust domain.

\subsubsection{Example Scenario}
Consider a logistics company with multiple regional warehouses. Each warehouse runs a set of IoT devices monitoring inventory and environmental conditions. These devices are grouped into a cluster per warehouse. Although the devices do not exchange data with one another, they are managed by the same IT team and operate within the same secure network. By enforcing the a minimum client participation rule and monitoring selection behavior via a coordinator (cluster head), the system ensures robust protection against an adversarial aggregator that might try to isolate a single device for BSA.

\subsubsection{Cluster and Aggregator Communication}
\label{Cluster-aggregator-comm}

\begin{figure}
    \includegraphics[width=.9\linewidth]{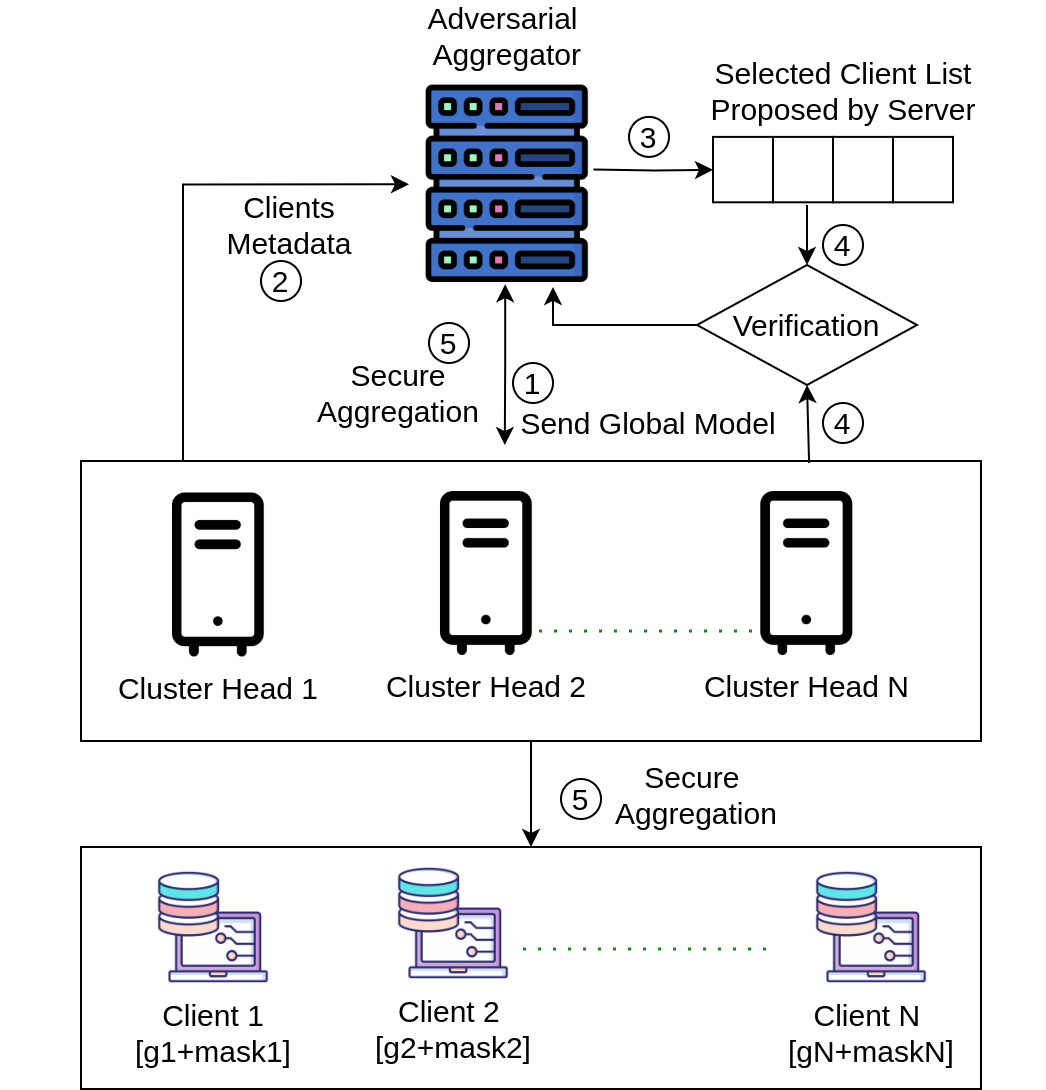}
    \caption{AdRo-FL for informed selection in cluster-oriented setting.}
    \label{cluster_oriented_FL}
\end{figure}

Fig.~\ref{cluster_oriented_FL} presents a high-level overview of the framework, illustrating the interaction flow between the aggregator, cluster heads, and clients. The communication and decision-making process are designed to ensure client privacy, prevent BSA. The process goes through in five main steps, as detailed below:

%\setlength{\itemindent}{-1em}
%\begin{enumerate}
\textit{Step 1) Global Model Broadcast}: The aggregator shares the latest global model with all cluster heads. Each cluster head, responsible for coordinating communication within its cluster, distributes this model to the local clients. This ensures that all clients work on a consistent starting point for their local training.

\textit{Step 2) Client Metadata Collection}: After receiving the global model, each client starts local training. Then it computes a function called client utility function (described in details in section \ref{utility_calculate}. The cluster head shares this utility value along with some metadata (client IDs, cluster IDs, payload transmission time) with the aggregator. No raw model updates or private data are shared at this stage, maintaining client confidentiality.

\textit{Step 3) Client Selection by the aggregator}: Using the received utility values, the aggregator performs client selection for the upcoming FL round. Selection can be conducted \textit{cluster-wise} (i.e., locally), or globally (based on the global view of all clients utility values from all clusters). Clients are first filtered based on their ability to meet transmission deadlines (discussed in section \ref{Methodology_Efficient_Client_Selection}) and then sorted based on the utility values. The aggregator can limit how many clients can participate at each round and adjust the sorted list accordingly. This sorted list is shared with the cluster heads.

\textit{Step 4) Privacy Verification by Cluster Heads}: Once the aggregator publishes the list of selected clients, each cluster head reviews the list to ensure compliance with the privacy constraint, i.e., the minimum participation threshold ($C$ clients per cluster, as discussed in section \ref{prevent_bsa_in_cluster}). If a violation is detected by a cluster head, it withholds its cluster members participation for that round, preventing the aggregator from isolating individual clients to do BSA.

\textit{Step 5) Secure Aggregation Participation}: Finally, the selected and approved clients participate in the SA protocol. Clients send their masked model updates to the aggregator. This completes one round of FL while maintaining client privacy and ensuring resilience against biased selection attack.
%\end{enumerate}

The whole process of client selection for cluster-oriented client setting is formalized in algorithm \ref{alg:cluster_oriented_selection}.

\begin{algorithm}[H]
\caption{Secure, Informed Client Selection in Cluster-oriented Client Setting}
\label{alg:cluster_oriented_selection}
\begin{algorithmic}[1]
    \Statex \textbf{Input:} Global model \(\theta\), clusters \(J\), \(n_j\) means clients in cluster $j$ where $j \in J$, transmission deadline \(D\), privacy threshold \(C\), client utility function $\mathcal{H}(\omega)$, global sample size \(\mathcal{S}_{\text{global}}\), local sample size \(\mathcal{S}_{\text{local}}\), client selection scope E
    \Statex \textbf{Output:} Participating clients for secure aggregation (SA)
    
    % Step 1: Global Model Broadcast
    \State \textbf{Global Model Broadcast:}
    \State Aggregator sends global model \(\theta\) to all cluster heads
    \For{each cluster head \(j \in J\)}
        \State Distribute \(\theta\) to all clients \(i \in n_j\)
    \EndFor
    
    % Step 2: Client Metadata Collection
    \State \textbf{Client Utility Collection:}
    \For{each client \(i \in n_j\) for each \(j \in J\)}
        \State Compute local training with \(\theta\)
        \State Compute transmission time \(T_{i,j}\), Client Utility $\mathcal{H}(\omega)$
        \State Send \(M_i = (t_i, \mathcal{H}(\omega) )\) to cluster head \(j\)
    \EndFor
    \For{each cluster head \(j \in J\)}
        \State Send \(\{M_i\}_{i \in n_j}\) to aggregator
    \EndFor
    
    % Step 3: Client Selection by Aggregator
    \State \textbf{Client Selection by Aggregator:}
    \State Initialize selected clients set \(S = \emptyset\)
    \For{each cluster \(j \in J\)}
        \State Let \(F_j = \{i \in n_j \mid T_{i,j} \leq D\}\) \Comment{Filter by transmission time}
        \State Sort \(F_j\) by $\mathcal{H}(\omega)$ in descending order either locally or globally based on $E$: let \(A_j\) be the sorted list
        \State Select a subset \(S_j \subseteq A_j\) based on ranking (e.g., top-\(L\) or all clients meeting a threshold)
        \State Add \(S_j\) to \(S\): \(S = S \cup S_j\)
    \EndFor
    \State Aggregator publishes selected clients list \(S\)
    
    % Step 4: Privacy Verification by Cluster Heads
    \State \textbf{Privacy Verification by Cluster Heads:}
    \For{each cluster head \(j \in J\)}
        \If{\(|S \cap n_j| \geq C\)}
            \State Let participating clients from cluster \(j\) be \(W_j = S \cap n_j\)
        \Else
            \State Let participating clients from cluster \(j\) be \(W_j = \emptyset\) \Comment{Withhold participation}
        \EndIf
    \EndFor
    \State Let final participating clients be \(W = \bigcup_{j \in J} W_j\)
    
    % Step 5: Secure Aggregation Participation
    \State \textbf{SA Participation:}
    \For{each client \(i \in W\)}
        \State Client \(i\) sends masked model update \(x_i\) to aggregator through SA
    \EndFor
    \State Aggregator updates global model  \(\theta^{t+1} \), then loops to step1
\end{algorithmic}
\end{algorithm}

\subsubsection{Preventing BSA in Cluster-oriented Setting}
\label{prevent_bsa_in_cluster}
Here we describes how AdRo-FL prevents BSA shown in the threat model.

\begin{figure*}
\centering
\includegraphics[width=\textwidth]{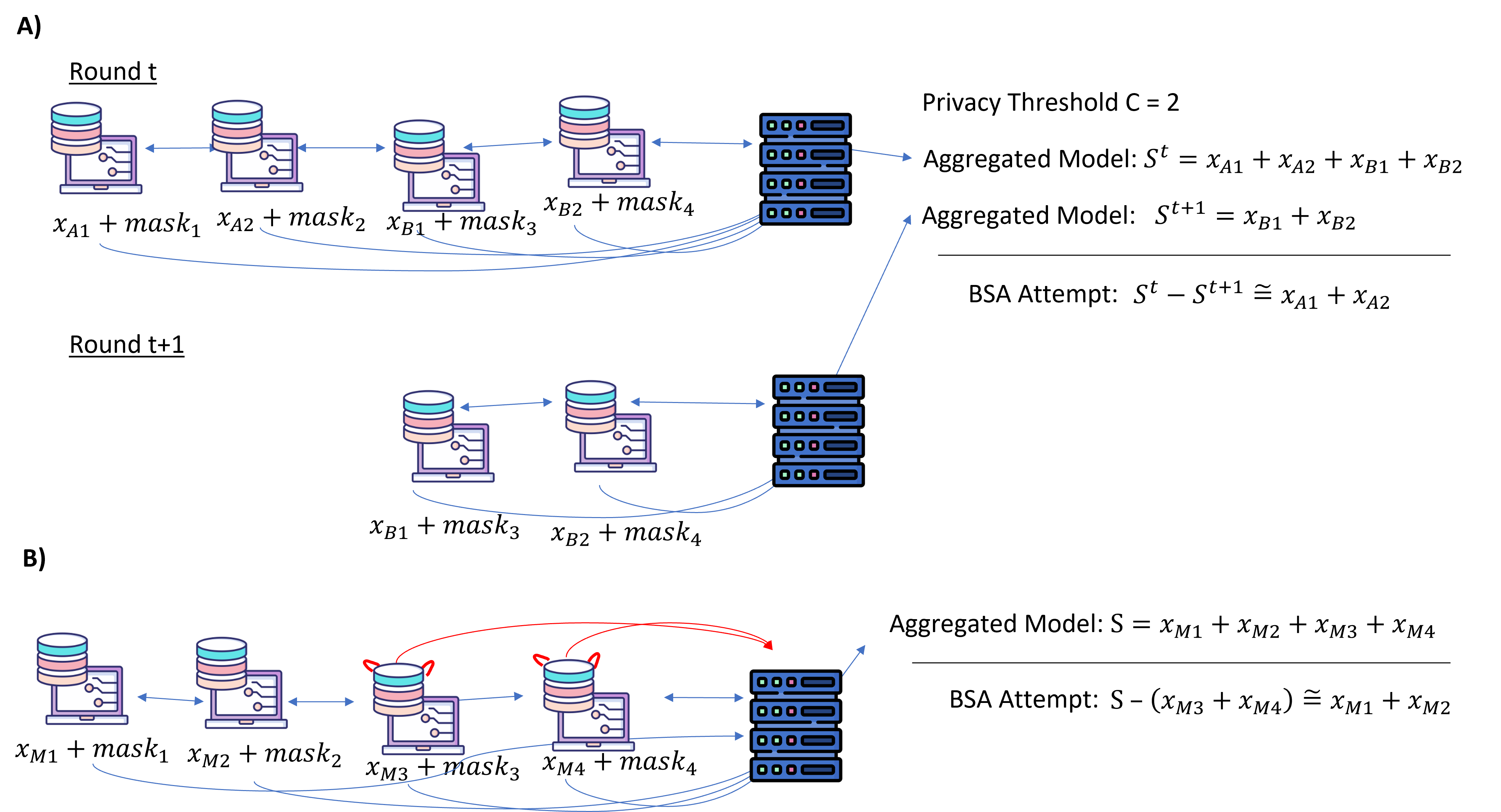}
\caption{A) BSA attempt for non-colluding client setting. B) BSA attempt for colluding client setting.}
\label{bsa_clusterFed}
\end{figure*}

Although the BSA mentioned in section \ref{threatmodel} did not consider a cluster-oriented client setting, it can very well be applied on it. To prevent such attack on cluster-oriented client setting, we assume clients are grouped into clusters with trust limited to within each cluster or at least a minority of the clients are trusted in the cluster. We introduce a privacy threshold (\(C\)). It imposes that each cluster contributes at least \(C\) clients or none. This threshold ensures that a malicious aggregator receives aggregated updates from at least \(C\) clients from a cluster. Though setting \(C = 2\) ensures that the aggregator cannot obtain a single client's model update, setting \(C > 2\) further strengthens security by increasing client resistance to BSA. Without this restriction on clusters, a single client is free to join from a cluster in SA making it vulnerable to BSA. This constraint ensures the privacy of each client by mandating that only clusters with a selected client count greater than or equal to \( C \) submit their member client models to the aggregator, while clusters with fewer than \( C \) selected clients withhold their participation.

Now we give an example on how AdRo-FL prevents a non-colluding BSA by using similar attack scenario mentioned section \ref{threatmodel}. If \(C=2\), there will be either at least 2 clients from each contributing cluster or none. At round \(t\), the malicious aggregator tries to select a victim with a subset of other clients for SA. Assuming there are four clients from two different clusters (Cluster \(A\) and \(B\)). With \(C=2\), the aggregator obtains, for instance, \(S^t=x_{A1}+x_{A2}+x_{B1}+x_{B2}\) through SA. Here, $x_{A1}$ denotes the model update from client $A1$ and so on. At round \(t+1\), the aggregator attempts to re-select the same subset of clients excluding a victim client (for instance, \(x_{A1}\)). Here the malicious aggregator wants to select just one client (\(x_{A2}\)) from cluster \(A\). However, since the privacy threshold prevents the cluster \(A\) from donating less than $C=2$ clients, it will withhold from contributing any clients on round \(t+1\). Hence, the aggregator obtains \(S^{t+1}=x_{B1}+x_{B2}\) where no clients participate from Cluster A. Therefore, the aggregator fails to uncover \(x_{A1}\) by estimating \(S^t - S^{t+1}\).

As for colluding BSA, AdRo-FL prevents it in a similar way as discussed above. The aggregator attempts to select a victim client \( V \) and a subset of colluding clients. Assume \( M3 \) and \( M4 \) are two colluding clients (or Sybils), and \( M1 \) and \( M2 \) are two honest clients from a cluster. However, the colluding clients (\( M3, M4 \)) secretly share their model updates with the aggregator. Subsequently, the aggregator, via an SA protocol, obtains \( S = x_{M1} + x_{M2} + x_{M3} + x_{M4} \). Since the aggregator knows the local models \(x_{M3}\) and \(x_{M4}\) of the colluding clients, it attempts to approximate the victim’s model as \( x_V = S - (x_{M3} + x_{M4}) \). This attempt fails because there will be at least \(C\) clients aggregated model updates in estimated \(x_V\), not single model update (assuming \(C=2\)). Hence the aggregator will obtain \( x_V \approxeq x_{M1} + x_{M2} \).

Furthermore, this scheme protects the privacy of other clusters in case a fake cluster is created by the aggregator. The aggregator may attempt to create colluding clients from the fake cluster but will fail to draw fewer than \( C \) clients from other real clusters due to the privacy threshold.

This is to be noted that the defense for cluster-oriented setting is safe against non-colluding attack, as we showed above. As for colluding attacks, there are some nuances. If the aggregator creates a fake cluster or fake clients, AdRo-FL is safe, as the real cluster will still follow the \(C\) threshold for client participation which prevents BSA. However, it becomes difficult to prevent BSA on a cluster when there is colluding client(s) within the cluster. For instance, if \( C \) is set to 2 and one client from a cluster colludes with the aggregator, then the aggregator will be able to target a client from that cluster. To prevent BSA in such case, we need to set \(C\) to at least 3. Similarly, if two clients from a cluster colludes with the aggregator, then \(C\) should be set to at least 4 and so on. Therefore, in such cases, we suggest to set the value of \(C\) to $\geq t+2$ where $t$ is the number of clients in a cluster that may collude with the aggregator. 

However, in many real-world deployments, we do not know the exact number of colluding clients in a cluster. Instead, we assume that each client has an independent probability \( \phi \) of colluding with the aggregator. To protect against BSA, we propose a minimum client selection threshold \( C \) such that the number of honest clients among the selected group is at least 2 with high probability. Let \( X \sim \text{Binomial}(C, 1 - \phi) \) denote the number of honest clients among the \( C \) selected clients. Our goal is to ensure that the probability \( P(X < 2) \leq \delta \), where \( \delta \) is a small risk tolerance parameter (e.g., \( \delta = 0.01 \)). This condition can be satisfied by solving for the smallest \( C \) such that \( \phi^C + C \cdot \phi^{C-1}(1 - \phi) \leq \delta \). For example, if \( \phi = 0.3 \) and \( \delta = 0.01 \), then the required value of \( C \) is 12.

Since this calculation may be required frequently in systems with varying levels of assumed collusion or changing security requirements, we recommend precomputing a table of minimum required \( C \) values for common combinations of \( \phi \) and \( \delta \). This avoids runtime computation and enables fast, adaptive policy enforcement by simply looking up the appropriate threshold from the table. We provide a sample precomputed lookup table for selecting an appropriate \( C \) in Appendix~\ref{appendix:min-c-table}.

\subsection{\textbf{Informed client selection for distributed, non-cluster-oriented clients}}
\label{informed_selection_VRF}

In this section we describe how informed client selection will be achieved when clients are distributed and there are no trust among clients nor there are any cluster-oriented client setting. Here AdRo-FL does secured informed client selection in two levels. Below we describe the process in details.

\subsubsection{First level of selection}
\label{vrf_first_level_selection}
\begin{figure}[htbp]
\centering
\includegraphics[width=\columnwidth]{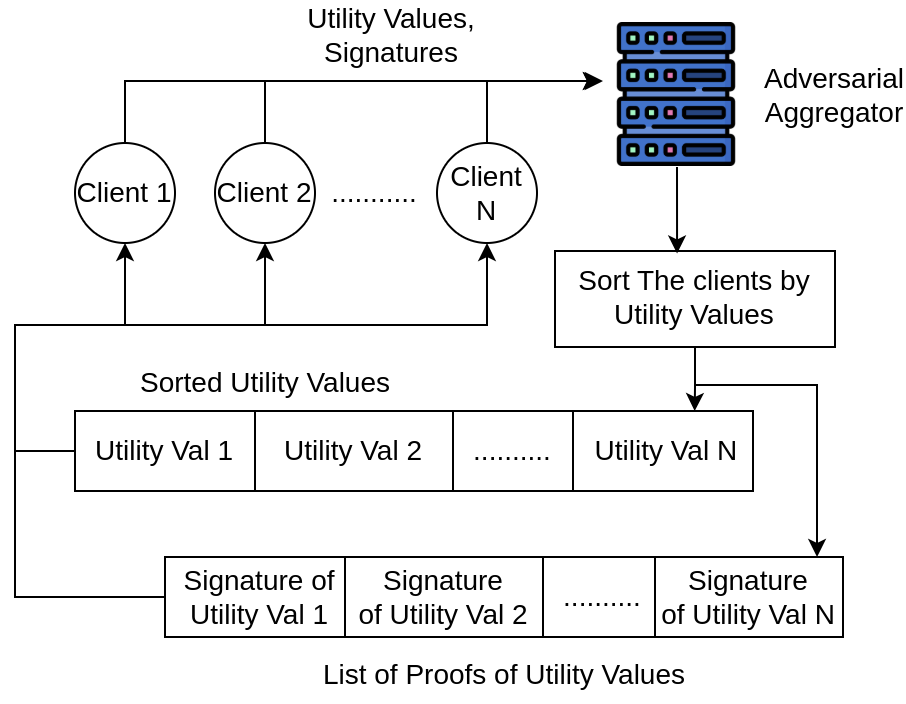}
\caption{AdRo-FL: First level of filtering for informed client selection in distributed, non-cluster-oriented setting.
}
\label{first_level_selection}
\end{figure}

At the first level, after local training, clients are filtered based on a predefined transmission deadline \(D\). The surviving clients compute their local utility value using equation \ref{eq:hybrid_metric}. They also produce signatures (like digital signature) for this utility values using their private keys, and submit both the raw utility values and the signatures to the aggregator. The aggregator first ranks these utility values from highest to lowest and saves them into a list. Then it maps the corresponding signatures in another list in the same order and compress it for size reduction. The aggregator broadcasts these two lists for all clients. If a client positions in the top 80\% of the sorted utility list, it will participate in the secondary selection process. Any client can verify if the aggregator altered any client's utility value by using the lists and corresponding client's public key. Since every client can explicitly verify the authenticity of the entire selection, all clients can be certain that their selection or non-selection in the first level is completely uninfluenced by the aggregator. In fig \ref{first_level_selection}, we illustrate this as how the primary selection is done.

As for how each client generates the signature and how it is utilized for verification, we propose the following method. Each client digitally signs its numeric utility value using its private key (for e.g., using Ed25519~\cite{brendel2021provable}). The aggregator collects the utility values and signatures from all clients, concatenating utility values into a compact binary string and signatures into fixed-sized indexed chunks where each chunk is independently compressed (for e.g., using optimal-level (`zlib') compression~\cite{maulidina2023comparative}). The server then publicly broadcasts these compressed signature chunks along with the concatenated utility values. Upon reception, each client identifies and decompresses only the relevant chunk containing the desired signature, precisely extracts the numeric value and signature based on client indices, and verifies the signature using the corresponding client’s public key. This compression and indexing approach significantly reduces payload sizes to enable efficient signature verification suitable for resource-constrained IoT environments. However, more efficient techniques can be used to further reduce the payload size depending on the target IoT devices.

\subsubsection{Second level of selection}
\label{vrf_second_level_selection}
\begin{figure}[htb!]
    \centering
    \includegraphics[width=\linewidth]{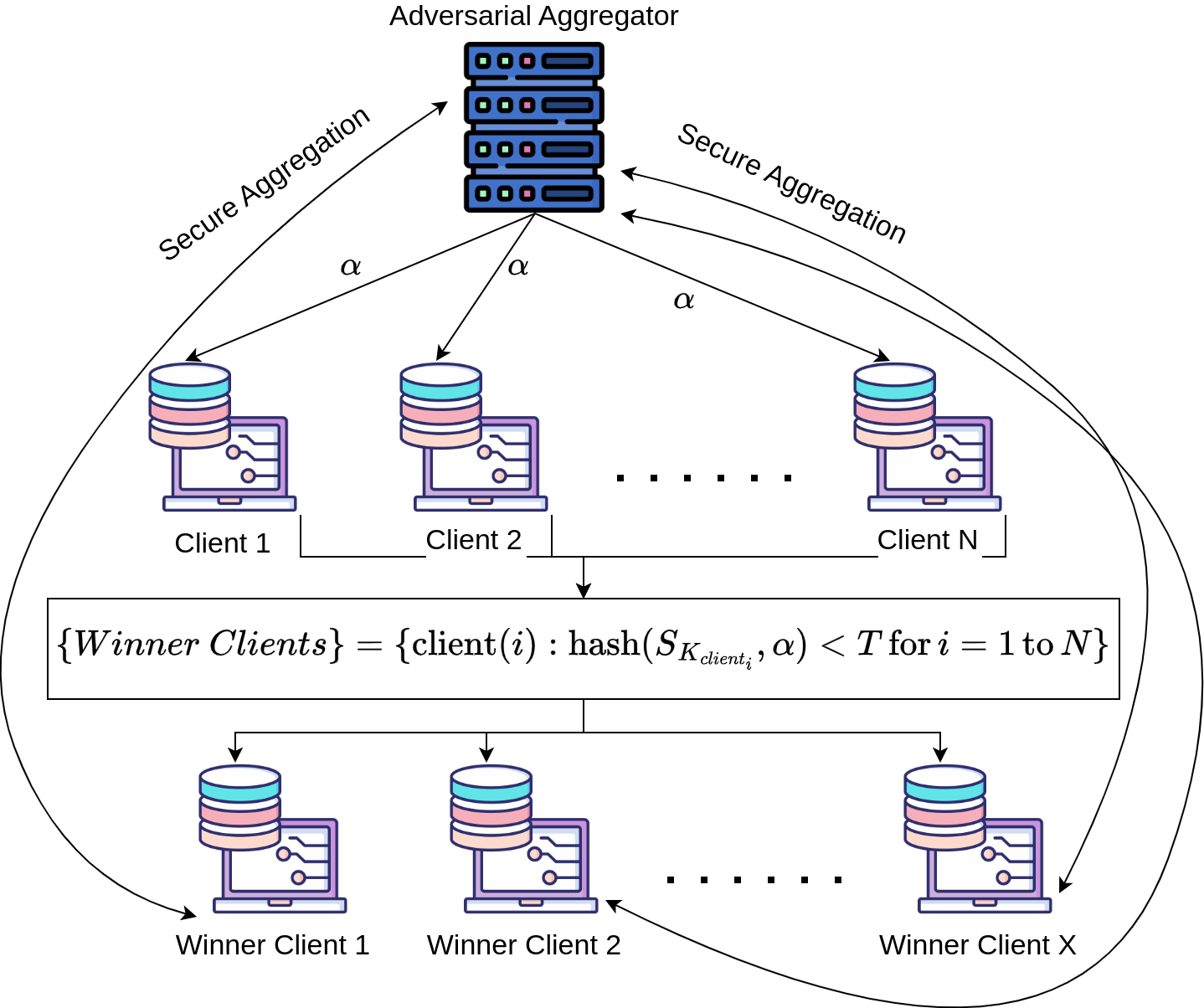}
    \caption{
    AdRo-FL: Second level of filtering with verifiable random client selection in distributed, non-cluster-oriented setting.}
    \label{fig:vrf_second_level}
\end{figure}

To achieve verifiable randomness, we assume that clients use some form of lightweight cryptography to generate private and public keys. In Fig. \ref{fig:vrf_second_level}, we illustrate the second-level client filtering process, which consists of three key steps:

\textit{Step 1) Generating VRF \( \alpha \):} Here, the top 80\% values in the sorted utility list from the first level of selection will be concatenated into a string and will be used as the \( \alpha \) value for VRF. The reason we don't use a simple deterministic value for VRF \( \alpha \) (for e.g., FL round number) is that this will give the aggregator more advantage to create Sybils. For instance, if the aggregator know what will be the \( \alpha \) in future rounds, it can carefully generate public key, private key pairs for Sybils that satisfies the VRF threshold-based selection. The aggregator can compute Sybil's hash using $VRF(Sk, \alpha)$, then check if the hash is less than the threshold and repeat until it satisfies.

\textit{Step 2) Generation of VRF Hash and VRF Proof:} The \( \alpha \) value from the previous step is used as input to the VRF to generate a new value called \( \beta\). This \( \beta\) is a cryptographic hash, created using secure hash functions like SHA-256 or SHA-512. Along with \( \beta \), the VRF also produces a proof \( \pi \). This proof acts as a cryptographic guarantee that \( \beta \) was generated correctly. It is created using either RSA or elliptic curve cryptography (ECC)~\cite{papadopoulos2017making}. Since clients utilize their private keys to independently generate unique hash values and corresponding proofs, both the authenticity and uniqueness of the generated values are ensured.

\textit{Step 3) Selection of Winner Clients:} Each client compares its VRF hash \( \beta \) against a threshold \(Z\). If the hash value is less than \(Z\), the client is selected for participation in the upcoming FL round. The threshold \(Z\) controls how many clients are chosen. Let's assume the hash uses 512-bit. If \(Z = 2^{512}\), all clients will be selected; if \(Z = 0\), none are selected. This is because the \( \beta \) will be compared against $Z$. Thus, \(Z\) directly influences client participation. Let's assume that in order to eliminate the risk of BSA, we need to select at least \(K\) clients out of total clients \(N\). Choosing proper value of \(K\) ensures that the chance of a BSA remains negligible. First, we determine the minimum \(K\) needed for a desired attack probability \(P\). Then, we calculate the threshold \(Z\) based on \(K\) and the total number of clients \(N\). A biased selection attack happens when a dishonest aggregator selects mostly colluding clients and isolates a single honest one (discussed in section \ref{threatmodel}). Lets assume \(X\%\) of \(N\) clients are honest and \(Y\%\) are colluding. The aggregator randomly chooses \(K\) clients from $N$ (enforced by AdRo-FL using VRF). The probability \(P\) of selecting exactly one honest and \(K - 1\) colluding clients is given by \(P = \frac{\binom{N \cdot X\%}{1} \binom{N \cdot Y\%}{K - 1}}{\binom{N}{K}}\). To ensure that this attack probability stays below a safe threshold, we require that \(\frac{\binom{N \cdot X\%}{1} \binom{N \cdot Y\%}{K - 1}}{\binom{N}{K}} \leq P\). 
    
Solving this inequality gives the value of \(K\) needed to satisfy the desired security level. Once \(K\) is known, we estimate the threshold $Z$ using \(K = N \times \frac{Z}{2^{512}}\), which reflects that the probability of a hash being below \(Z\) is proportional to the ratio \(Z / 2^{512}\). Rearranging gives \(Z = \frac{K \times 2^{512}}{N}\). However, due to randomness, fewer than \(K\) clients may be selected in some rounds upon applying the threshold-based selection. To address this, we recommend multiplying the term by a conservative factor \(f\) whose value can be 2 or 3 depending on the use case. Hence the general formula to find threshold $Z$ becomes: $ Z = \frac{f \times K \times 2^{512}}{N} \label{eq:formula_Z}$. Clients whose VRF hash values are lower than \(Z\) belong to the winning client set \(W\):
    \[
    \{W\} = \{ \text{client}(i) : \text{hash}(client_i) < Z \, \text{for} \, i = 1 \, \text{to} \, N \}
    \]
These clients participate in the secure aggregation process. To ensure fairness, any client can verify another client's selection by checking the VRF hash, proof \(\pi\), and public key. Clients not in \(\{W\}\) do not send updates. This two level client selection approach enables informed, unbiased and verifiable random client selection in a distributed setting.

The whole process of client selection for non-cluster-oriented client setting is formalized in algorithm \ref{alg:non_cluster_selection}. In the following section we discuss the security aspects of this framework.

\begin{algorithm}[t]
\caption{Secure, Informed Client Selection in Non-Cluster-oriented Client Setting}
\label{alg:non_cluster_selection}
\begin{algorithmic}[1]
    \Statex \textbf{Input:} Global model \(\theta\), set of all clients \(N\), transmission deadline \(D\), conservative factor \(f\), client utility function \(\mathcal{H}(\omega)\), client's secret key $sk_i$, VRF hash Threshold \(Z\) for client selection
    \Statex \textbf{Output:} Participating clients for secure aggregation (SA)

    % Step 1: Global Model Broadcast
    \State \textbf{Global Model Broadcast:}
    \State Aggregator sends global model \(\theta\) to all clients \(i \in N\)

    % Step 2: First-Level Selection
    \State \textbf{First-Level Selection:}
    \For{each client \(i \in N\)}
        \State Evaluate transmission time \(T_i\)
        \If{\(T_i \leq D\)}
            \State Compute local client utility score \(\mathcal{H}(\omega_i)\)
            \State Generate signature \(\sigma_i = \text{Sign}(\mathcal{H}_i)\)
            \State Submit \((\mathcal{H}_i, \sigma_i)\) to aggregator
        \EndIf
    \EndFor
    \State Aggregator ranks clients by utility \(\mathcal{H}(\omega_i)\), compresses the signature into a list and publishes the ordered lists of utilities and corresponding signatures
    \For{each client \(i\)}
        \If{\(i\) is in top \(80\%\) of sorted utility list}
            \State Mark client \(i\) eligible for second-level selection
        \EndIf
        \State Verify \((\mathcal{H}_q, \sigma_q)\) for all \(q \in N\) using public keys to ensure authenticity of the lists \Comment{Optional step}
    \EndFor

    % Step 3: Second-Level Selection Using VRF
    \State \textbf{Second-Level Selection Using VRF:}
    \State Concatenate top \(80\%\) utility scores into a string to form VRF input \(\alpha\)
    \For{each eligible client \(i\)}
        \State Generate VRF hash \(\beta_i = \text{VRF}(\text{sk}_i, \alpha)\)
        \State Generate proof \(\pi_i\) to accompany \(\beta_i\)
    \EndFor
    \State \textbf{Winning Client Selection:}
    \State Initialize \(P \leftarrow \emptyset\)
    \For{each client \(i\)}
        \If{\(\beta_i < Z\)}
            \State Add client \(i\) to \(P\)
        \EndIf
    \EndFor

    % Step 4: Secure Aggregation Participation
    \State \textbf{SA Participation:}
    \For{each client \(i \in P\)}
        \State Client \(i\) sends masked model update \(x_i\) to aggregator through SA
    \EndFor
    \State Aggregator updates global model  \(\theta^{t+1}\), then loops to step1

\end{algorithmic}
\end{algorithm}

\subsubsection{Security Properties Analysis}
AdRo-FL withstands the following security challenges:\\

\textit{Pool Consistency:} Pool consistency ensures all clients observe the same selection outcome, preventing the aggregator from showing different selection results to different clients. In our approach, clients are selected using a VRF hash and a public threshold \(Z\), making the process both random and verifiable. Any client can independently verify the selected set by checking the VRF  \( \alpha \), proofs \( \pi \), hash $\beta$, and public keys. This verification ensures transparency and prevents manipulation by the aggregator.

\textit{Pool Quality:} Pool quality requires that the final selected set includes a minimum number of honest clients. A dishonest aggregator may try to favor colluding clients. To counter this, we ensure the probability \(P\) of selecting exactly one honest and \(K-1\) colluding clients remains negligible. This is done by tuning \(K\) such that the chance of such biased selection is statistically insignificant. As a result, the final pool \(\{W\}\) reliably includes a sufficient number of honest participants.

\textit{Anti-Targeting:} Anti-targeting prevents the aggregator from selectively including or excluding specific clients. Since the aggregator does not know the VRF  \( \alpha \) or hash $\beta$ in advance, it cannot influence the selection outcome. VRF outputs, derived using SHA-512, are uniformly random. Thus, every honest client has an equal chance of being selected, with the probability of selection proportional to \(Z / 2^{512}\). This guarantees fairness and prevents targeted manipulation.

\section{\textnormal{AdRo-FL}: Efficient Client Selection}
\label{Methodology_Efficient_Client_Selection}
\subsection{Federated Learning Optimization}
\label{optimization}
We adopt Federated Stochastic Gradient Descent (FedSGD)~\cite{mcmahan2017communication} as the foundation for collaborative learning in AdRo-FL. FedSGD is selected over other FL optimization approaches such as FedAvg as it takes less local computational time per round, making it particularly suited for scenarios with high computational heterogeneity. Besides, it is typically more memory-efficient as clients process fewer batches per round. FedSGD enables multiple distributed clients to jointly train a shared global model while preserving data privacy. Instead of transmitting raw data, clients compute and transmit model gradients, ensuring that sensitive information remains on local devices. In each iteration of FedSGD, the aggregator first distributes the latest global model to all participating clients. Each client then computes the stochastic gradient of its local objective function based on the received global model. These locally computed stochastic gradients are subsequently transmitted to the aggregator, which aggregates them and applies a global SGD update step. This iterative process continues until the model reaches convergence. 
%Following is a mathematical description of the client-aggregator computations at iteration $t+1$. 
Each client computes its local gradient at round \( t \): $g_i^t = \nabla f_i(\theta^t)$. The aggregator aggregates these updates and computes the global parameters for round \( t+1 \): $\theta^{t+1} = \theta^t - \alpha \sum g_i^t$, where \(\theta^t\) denotes the global model parameters at communication round \(t\), and \(\alpha\) represents the learning rate. However, we introduce an additional optimization layer aimed at minimizing overall energy consumption and accelerating convergence at the aggregator. It also considers imposing a minimum client requirement per cluster for cluster-oriented client setting. The optimization formulation for AdRo-FL is defined as follows:

\begin{align}
\underset{I, \theta}{\textbf{min}} &\Big[\sum_{j=1}^J\sum_{i \in n_j} (f_{i,j}(\theta) - \sum_{t=1}^{t_{max}} \sum_{j=1}^J\sum_{i \in n_j} I_{i,j}^t \cdot \| \nabla f_{i,j}(\theta^t) \|_2)\Big]\label{formulation}\\
\text{subject to} \nonumber \\
&R_{i,j}^t = B_{i,j}\log_2(1+SNR_{i,j}^t)\\
&T_{i,j}^t = I_{i,j}^t \cdot \frac{PL_q}{R_{i,j}^t} \label{transTime} \\
&0 \leq T_{i,j}^t \leq D \label{deadline} \\
& I_{i,j}^t \in \{0,1\}\\
&\sum_{i \in n_j} I_{i,j}^t \in \{0, \ge C\}, \forall j\label{threshold}
\end{align}

\begin{table}[]
\centering
\caption{Summary of Notation}
\label{notation-table}
\scriptsize % Reduces font size within the table
\begin{tabular}{|c|p{6.7cm}|}
\hline
\textbf{Notation} & \textbf{Explanation} \\ \hline
\(N\) & Total number of clients \\ \hline
\(J\) & Total number of clusters \\ \hline
\(t_{max}\) & Total number of federated learning rounds \\ \hline
\(n_j\) & Total clients belonging to cluster \(j\) \\ \hline
\(I_{i,j}^k\) & Scheduling decision indicator (1 if client \(i\) in cluster \(j\) is selected at round \(t\), else 0) \\ \hline
\(B_{i,j}\) & Channel bandwidth allocated to client \(i\) in cluster \(j\) \\ \hline
\(SNR_{i,j}^t\) & Signal-to-noise ratio of client \(i\) in cluster \(j\) at round \(t\) \\ \hline
\(R_{i,j}^t\) & Achievable data rate (bits/sec) for client \(i\) in cluster \(j\) at round \(t\) \\ \hline
\(PL_q\) & Quantized payload size (bits)\\ \hline
\(T_{i,j}^t\) & Transmission time required by client \(i\) in cluster \(j\) to transmit \(PL_q\) \\ \hline
\(D\) & Deadline, Maximum allowed transmission time per round \\ \hline
\(C\) & Privacy threshold, Minimum clients needed from each cluster \\ \hline
\(\alpha\) & Learning rate \\ \hline
\end{tabular}
\end{table}

The notation used in this formulation is summarized in Table~\ref{notation-table}. The objective function~\ref{formulation} implies that we are minimizing the global loss while ensuring fast convergence and minimum energy consumption. A formal convergence analysis is included in the Appendix~\ref{appendix:convergence}. By selecting the clients based on the loss and gradient (detailed in section \ref{utility_calculate}) while they can meet the deadline (constraint \eqref{deadline}) with quantized payload, AdRo-FL speeds up the convergence and avoid transmission failures due to violating the deadline. Constraint \eqref{threshold} is related to the cluster-oriented client setting which enforces that the number of scheduled clients from each cluster should be $\ge C$ or $0$. This constraint \(C\) was discussed in details in section \ref{prevent_bsa_in_cluster}. %This constraint is imposed to ensure the privacy of each client by mandating that only clusters with a selected client count greater than or equal to \( C \) submit their models to the aggregator, while clusters with fewer than \( C \) selected clients will withhold their models. 
Equation \ref{formulation} equally applies to the non-cluster-oriented setting except that the cluster constraints are ignored. For instance, we assume \(C\) to be $0$ in non-cluster-oriented client setting.

\subsection{Utility-based Client Selection}
\label{utility_calculate}
To accelerate convergence, AdRo-FL considers a client’s local loss and weighted gradient norm. In the existing literature, various methods have been used for client selection to improve the global model, notably those based on local loss~\cite{cho2022towards} or gradient norm~\cite{marnissi2024client}. Clients with higher gradient norms likely represent data that is harder to fit or provides more informative gradients for global model. On the other hand, by selecting clients with higher local loss, the global model gets updates from clients where it performs worst. However, exclusively relying on gradient norm for measuring a client's contribution towards global model convergence is inefficient as gradient norms are more reflective of how well the global model is currently performing on a client’s specific data distribution. If the client’s data is mostly from classes that the model is struggling with, its gradient norms may be higher, even if the data is not diverse. It was observed~\cite{10.5555/3666122.3669518,9833969} that the gradient norm was significantly higher for clients with imbalanced data than those with comparatively balanced data. 

\vspace{-0.3cm}
\begin{equation}\label{eq:hybrid_metric}
\mathcal{H}(\omega) = \omega\mathcal{L}_{\text{local}} + (1-\omega) \|\nabla f(\theta) \|_2 \cdot \frac{|\mathcal{S}_{\text{local}}|}{|\mathcal{S}_{\text{total}}|} 
\end{equation}
\vspace{-0.3cm}

In order to leverage the effectiveness of both gradient norm and local loss, AdRo-FL propose a client utility computation formula in equation ~\ref{eq:hybrid_metric}. The proposed utility function $\mathcal{H}(\omega)$ combines local loss and gradient information to optimize client selection. We empirically determine the value of $\omega$. The first term, $\mathcal{L}_{\text{local}}$, represents the negative log likelihood loss, while $\|\nabla f(\theta) \|_2$ denotes the Euclidean L2 norm of the client's gradients. The ratios $|\mathcal{S}_{\text{local}}|/|\mathcal{S}_{\text{total}}|$ represent the client's proportion of samples. The L2 norm term quantifies the magnitude of model gradient for client's local data, allowing us to prioritize clients with higher gradient norms. This component proves particularly effective for highly imbalanced datasets. Conversely, the $\mathcal{L}_{\text{local}}$ term identifies how well or poorly the model's predictions align with the actual data labels on that specific client and is especially useful for balanced datasets. $\mathcal{H}(\omega)$ achieves a balanced approach, ensuring model generalization across diverse data distributions while maintaining responsiveness to challenging or unique data.

\subsection{Quantization}
AdRo-FL uses a fixed quantization for model gradients while ensuring that it does not degrade model performance. The quantization error term \(E_q\) is modeled as a function of a fixed quantization level \(Q\): $E_q = \frac{\sigma_q}{Q}$, where \(\sigma_q\) is a scaling factor that controls the quantization error magnitude. The quantization level remains the same throughout all FL rounds, leading to a consistent quantization error.

\section{Experimental Results}
\label{ER}
\subsection{Dataset}
We used MNIST, FMNIST, SVHN, and CIFAR10 datasets to assess AdRo-FL. For all datasets, we used Dirichlet distribution with alpha 0.1 to use high heterogeneity. The privacy threshold \(C\) was set to 2 for all the experiments related to cluster-oriented client setting.

\vspace{-0.3cm}

\subsection{Baselines}
%We took random, uninformed client selection technique as our first baseline. It is conventionally used approach in FL. Our second baseline is state of the art informed client selection technique called Oort\cite{273723}. It is a highly optimized informed client selection technique.

We choose two baseline client selection methods in FL. The first is a conventional random, uninformed client selection approach, commonly used in FL. The second baseline is Oort\cite{273723}, a state-of-the-art informed client selection technique known for its performance.

\vspace{-0.3cm}

%\subsection{Client Selection Scope}
%For the experiments with cluster-oriented clients, we experimented both local, cluster-wise client selection and global client selection. 
%In sectors like healthcare, regulations may mandate collaborative model training across institutions to ensure standardized care and diagnostics. All clusters must participate to comply with such mandates. Moreover, in systems where clusters represent different operational units (e.g., regional branches of a company), the performance of the global model depends on insights from all units. Omitting any cluster could compromise the model's utility across the organization.

\subsection{Environment Setting}
\label{EnvSet} 
\subsubsection{Constraints for Energy Efficiency}
\label{energy_setting}
For all experiments simulating communication, bandwidth is set at 1 \(Mhz\), and power at 0.1 \(js^{-1}\). \(SNR\) is calculated in decibels with a random value in the range of 0 to 30, then converted from dB to linear scale. Energy consumption is computed as \( (\text{power} \times \text{transmission time}) \). These values are chosen arbitrarily and can be modified according to the use case. As for the quantization level, 8 bit quantization was used for all datasets only for AdRo-FL. For the client utility function in ~\ref{utility_calculate}, $\omega$ was set to 0.4 based on empirical study. We used deadline \(D=0.5\) seconds for MNIST, \(D=0.11\) seconds for FMNIST, \(D=0.14\) seconds for CIFAR10 and \(D=0.13\) seconds for SVHN dataset. These deadline values are chosen based on the average transmission time found for each dataset with the above setting. All the above values can be tuned to suit the target application.

\subsubsection{FL Setting}
We implemented AdRo-FL as well as the baselines in Python using Pytorch. In training, all runs used 3000 rounds. For aggregation, we used FedSGD ~\cite{FEKRI2023109285}. For training clients, batch size was set to 64. The learning rate for AdRo-FL and random selection was set to 0.01 while for Oort the learning rate was set to 0.1. These values were chosen after tuning for better performance. The total number of clients were 100 and there were 10 clusters in the cluster-oriented client setting. 20 clients were selected at each FL round. The distribution of clients across clusters was as follows: [12,8,10,11,9,5,15,10,4,16]. Data was distributed across clients using Dirichlet distribution with alpha set to 0.1 for high degree of heterogeneity.

\subsubsection{Model Setting}
As for the model, the MNIST dataset uses a feed-forward network with two ReLU-activated hidden layers of 256 and 128 neurons, and a 30\% dropout after the first layer to prevent over-fitting. The final layer, using log-softmax, outputs probabilities across 10 classes. For training with FMNIST dataset, we used a multi-layer perceptron with two hidden layers of 64 and 30 units, using ReLU activations and dropout for regularization. For CIFAR10 and SVHN, a CNN was used. It consists two convolutional layers with max pooling, followed by three fully connected layers with dropout regularization and He initialization, culminating in a log softmax output for multi-class classification.

\subsubsection{Adversarial Setting}
For cluster-oriented client setting, we set cluster privacy threshold $C=2$. As for non-cluster-oriented client setting, we set the percentage of honest clients to 90\%, the percentage of colluding clients to 10\% and set the conservative factor $f=1$. We set the maximum tolerable probability of biased selection attack to 0.001. All the above values can be tuned to suit the target application.

\subsubsection{Hardware}
The experiments utilized a AMD Ryzen Threadripper PRO 5995WX 64-core processor with 500 GB of RAM and an NVIDIA RTX A6000 GPU with 48 GB of VRAM.

\vspace{-0.3cm}
\subsection{Results}
In this section, we present the performance of AdRo-FL on benchmark datasets. We also compare it with baselines. As mentioned in section \ref{Methodology_overview}, we consider two selection approaches for two different client settings. However, for cluster-oriented environment, we further divide it into local, cluster-wise selection and global selection. In the following subsections, we present the experimental outcomes. We report time-to-accuracy as the number of communication rounds required to reach a specified test accuracy threshold.

\subsubsection{Cluster-oriented Client Setting}
\begin{figure}[H]
    \centering
    \includegraphics[width=\columnwidth]{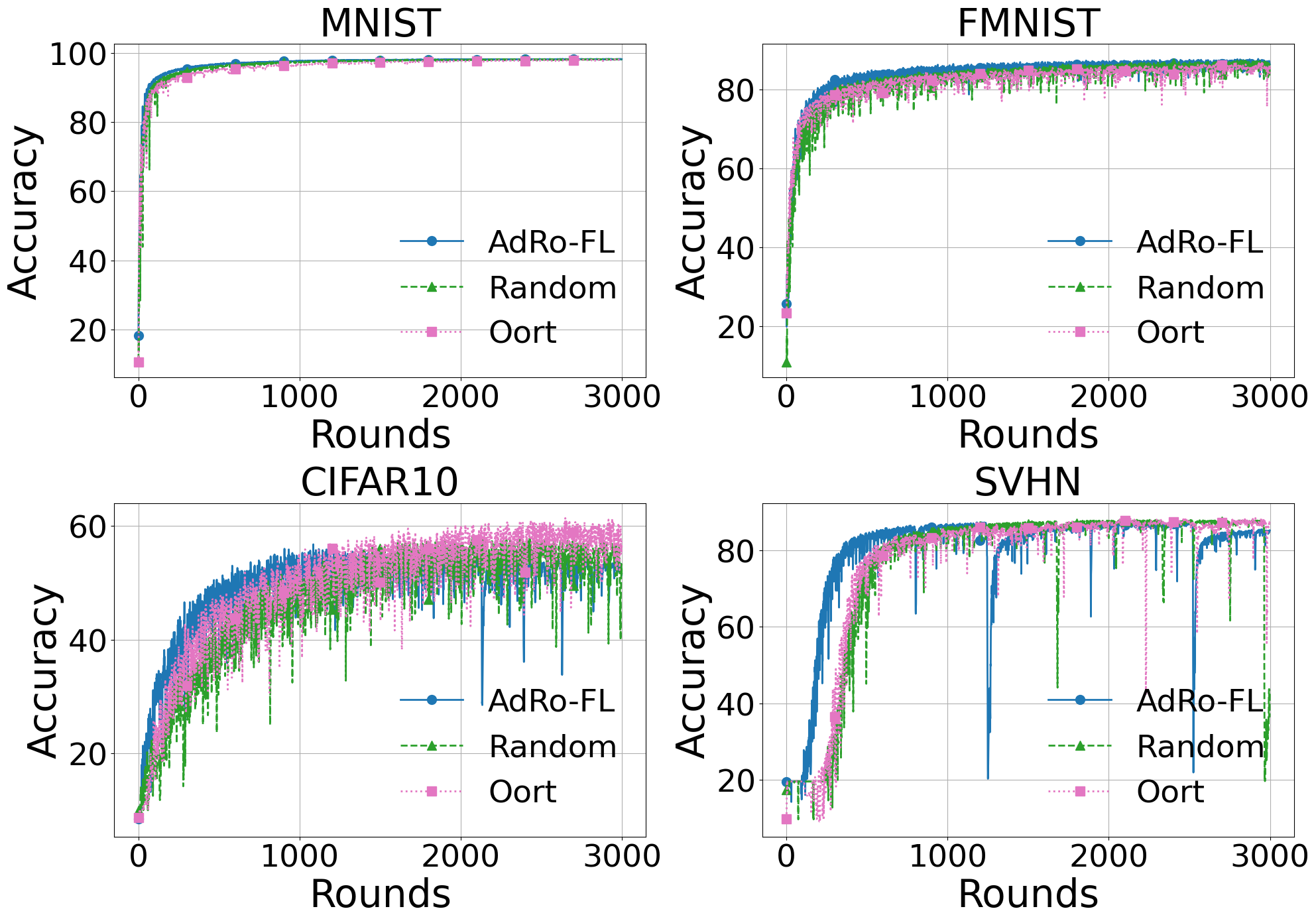}
    \caption{Accuracy comparison of AdRo-FL with random selection and Oort on four datasets. Here we used local selection (cluster-oriented).}
    \label{acc_comp_local_cluster}
\end{figure}

\begin{figure}[H]
    \centering
    \includegraphics[width=\columnwidth]{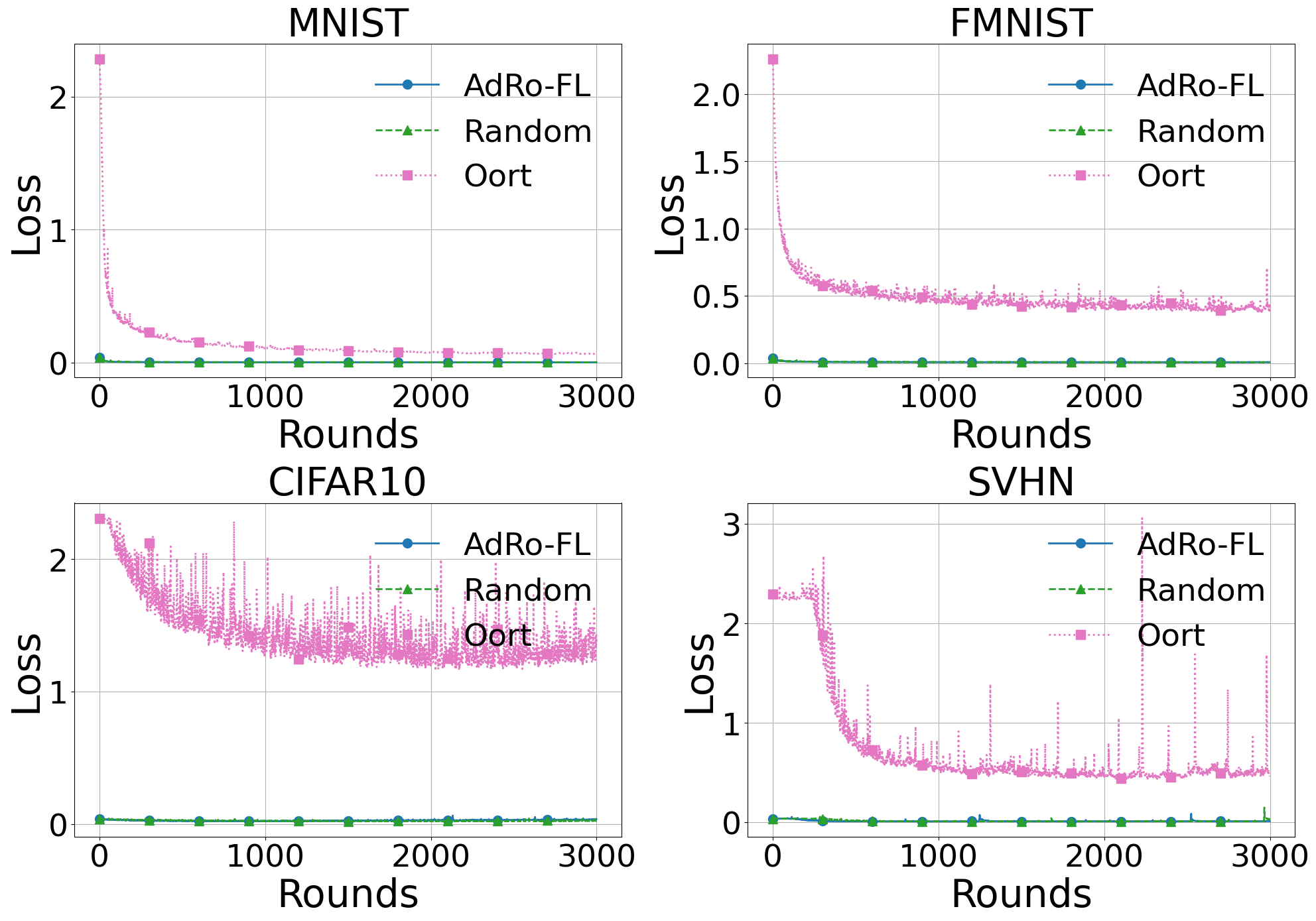}
    \caption{Loss comparison of AdRo-FL with random selection and Oort on four datasets. Here we used local selection (cluster-oriented).}
    \label{loss_comp_local_cluster}
\end{figure}

In this subsection, we report the performance of AdRo-FL. Besides, we present a comparison with the baselines, i.e., random selection and Oort. For each of the methods, we show the results of two approaches of selecting clients in cluster-oriented setting. First is local selection and the second is global selection. In local selection, we choose a fixed number of clients from each cluster while in global selection, clients are selected globally based on the view of all clients utility values. The random selection baseline method selects clients randomly while AdRo-FL and Oort selects client based on respective utility function. We present the results in Fig. ~\ref{acc_comp_local_cluster},~\ref{loss_comp_local_cluster},~\ref{acc_comp_local_cluster},~\ref{loss_comp_local_cluster}.

We start by local, cluster-wise selection. Fig. ~\ref{acc_comp_local_cluster} shows the accuracy of AdRo-FL and the baselines against the round numbers. The figures shows that AdRo-FL either achieves better or competitive performance compared to baselines. Fig. ~\ref{loss_comp_local_cluster} shows the loss values of AdRo-FL and the baselines against FL round numbers.

\begin{table}[H]
\caption{Time-to-accuracy and final test accuracy comparison on the MNIST dataset (Cluster-oriented setting). The best value for each row is highlighted in bold. Here clients were selected locally from all clusters.}
\label{tab:MNIST_local_cluster_results}
\centering
\resizebox{\columnwidth}{!}{%
\begin{tabular}{lcccccccc}
\hline
\textbf{Method} & \textbf{@ 60} & \textbf{@ 65} & \textbf{@ 70} & \textbf{@ 75} & \textbf{@ 80} & \textbf{@ 85} & \textbf{@ 90} & \textbf{Best Accuracy (\%)} \\
\hline
AdRo-FL & \textbf{12} & 20 & 20 & \textbf{21} & \textbf{27} & \textbf{40} & \textbf{63} & \textbf{98.23} \\
Random   & 16 & 18 & 23 & 23 & 37 & 54 & 86 & 98.20 \\
Oort     & \textbf{11} & \textbf{15} & \textbf{15} & 30 & 31 & 48 & 108 & 98.04 \\
\hline
\end{tabular}%
}
\end{table}

In table \ref{tab:MNIST_local_cluster_results}, we show the number of rounds needed to reach different target accuracies for MNIST dataset. Also we note the final accuracy achieved by each method. On average, compared to random baseline, AdRo-FL achieves $1.22\times$ time-to-accuracy improvement. Compared to Oort, it achieves $1.13\times$ time-to-accuracy improvement on average.

\begin{table}[H]
\caption{Time-to-accuracy and final test accuracy comparison on the FMNIST dataset (Cluster-oriented setting). The best value for each row is highlighted in bold. Here clients were selected locally from all clusters.}
\label{tab:FMNIST_local_cluster_results}
\centering
\resizebox{\columnwidth}{!}{%
\begin{tabular}{lccccccc}
\hline
\textbf{Method} & \textbf{@ 60} & \textbf{@ 65} & \textbf{@ 70} & \textbf{@ 75} & \textbf{@ 80} & \textbf{@ 85} & \textbf{Best Accuracy (\%)} \\
\hline
AdRo-FL & 43 & 52 & \textbf{56} & 116 & \textbf{195} & \textbf{639} & \textbf{87.75} \\
Random   & 50 & 57 & 89 & 156 & 311 & 1127 & 87.38 \\
Oort     & \textbf{39} & \textbf{43} & 69 & \textbf{114} & 312 & 1352 & 86.95 \\
\hline
\end{tabular}%
}
\end{table}

In table \ref{tab:FMNIST_local_cluster_results}, we show the number of rounds needed to reach different target accuracies for FMNIST dataset. Also we note the final accuracy achieved by each method (proposed and Oort). On average, compared to random baseline, AdRo-FL achieves $1.43\times$ time-to-accuracy improvement. Compared to Oort, it achieves $1.28\times$ time-to-accuracy improvement on average and $1.01\times$ final accuracy improvement.

\begin{table}[H]
\caption{Time-to-accuracy and final test accuracy comparison on the CIFAR10 dataset (Cluster-oriented setting). The best value for each row is highlighted in bold. Here clients were selected locally from all clusters.}
\label{tab:CIFAR10_local_cluster_results}
\centering
\resizebox{\columnwidth}{!}{%
\begin{tabular}{lcccccc}
\hline
\textbf{Method} & \textbf{@ 35} & \textbf{@ 40} & \textbf{@ 45} & \textbf{@ 50} & \textbf{@ 55} & \textbf{Best Accuracy (\%)} \\
\hline
AdRo-FL & \textbf{155} & \textbf{209} & \textbf{286} & \textbf{469} & \textbf{906} & 56.79 \\
Random   & 330 & 416 & 548 & 806 & 1334 & 58.15 \\
Oort     & 208 & 289 & 393 & 660 & 997 & \textbf{61.40} \\
\hline
\end{tabular}%
}
\end{table}

In table \ref{tab:CIFAR10_local_cluster_results}, we show the number of rounds needed to reach different target accuracies for CIFAR10 dataset. Also we note the final accuracy achieved by each method (proposed and Oort). On average, compared to random baseline, AdRo-FL achieves $1.85\times$ time-to-accuracy improvement. Compared to Oort, it achieves $1.32\times$ time-to-accuracy improvement on average.

\begin{table}[H]
\caption{Time-to-accuracy and final test accuracy comparison on the SVHN dataset (Cluster-oriented setting). The best value for each row is highlighted in bold. Here clients were selected locally from all clusters.}
\label{tab:SVHN_local_cluster_results}
\centering
\resizebox{\columnwidth}{!}{%
\begin{tabular}{lccccccc}
\hline
\textbf{Method} & \textbf{@ 60} & \textbf{@ 65} & \textbf{@ 70} & \textbf{@ 75} & \textbf{@ 80} & \textbf{@ 85} & \textbf{Best Accuracy (\%)} \\
\hline
AdRo-FL & \textbf{202} & \textbf{228} & \textbf{238} & \textbf{281} & \textbf{345} & \textbf{561} & 87.33 \\
Random   & 388 & 425 & 435 & 477 & 572 & 878 & 88.09 \\
Oort     & 354 & 379 & 404 & 438 & 535 & 966 & \textbf{88.44} \\
\hline
\end{tabular}%
}
\end{table}

In table \ref{tab:SVHN_local_cluster_results}, we show the number of rounds needed to reach different target accuracies for SVHN dataset. Also we note the final accuracy achieved by each method (proposed and Oort). On average, compared to random baseline, AdRo-FL achieves $1.76\times$ time-to-accuracy improvement. Compared to Oort, it achieves $1.66\times$ time-to-accuracy improvement on average.

Now we present the results of global selection in cluster-oriented client setting. 

\begin{figure}[H]
    \centering
    \includegraphics[width=\columnwidth]{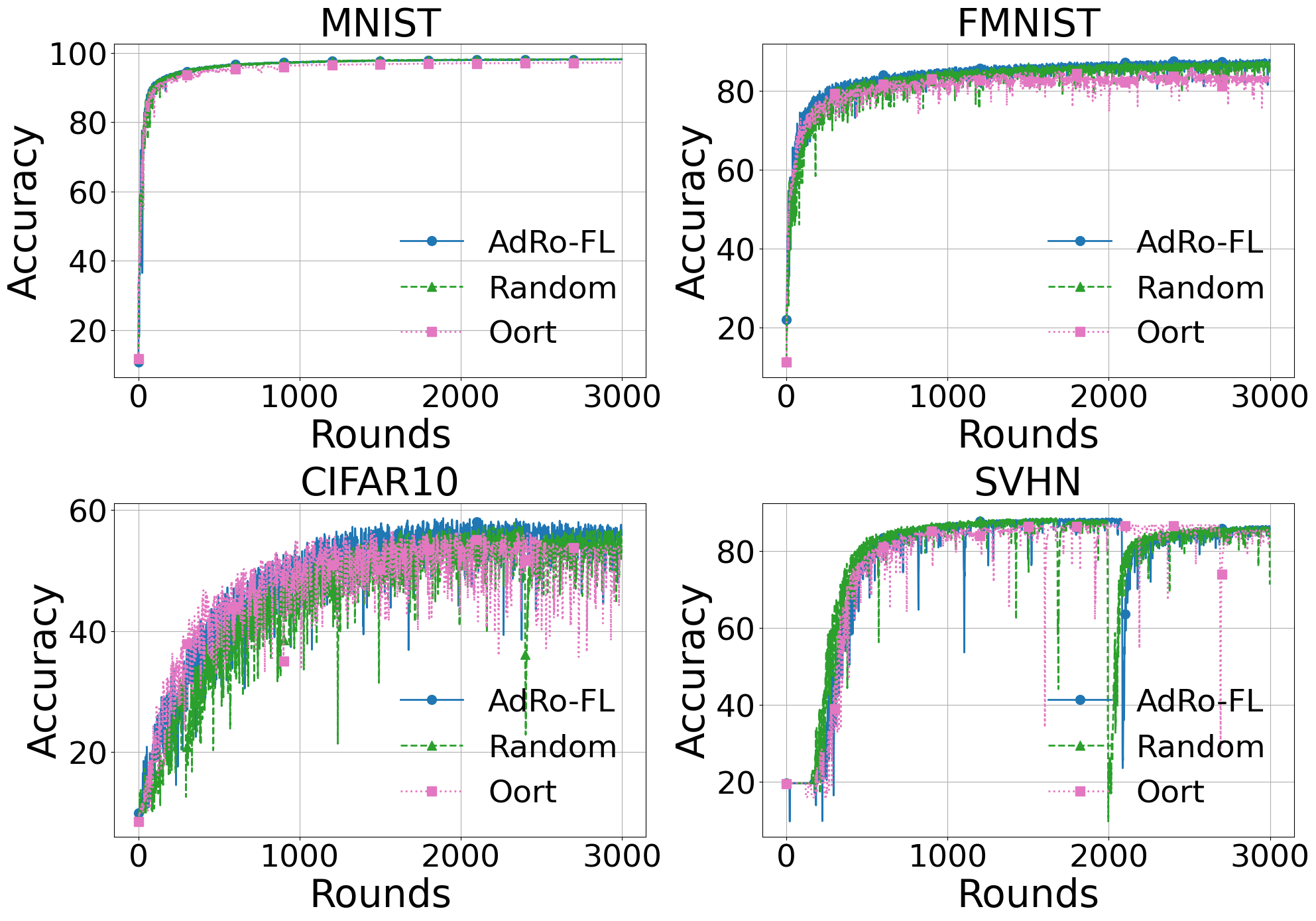}
    \caption{Accuracy comparison of AdRo-FL with random selection and Oort on four datasets. Here we used global selection (cluster-oriented).}
    \label{acc_comp_global_cluster}
\end{figure}

\begin{figure}[H]
    \centering
    \includegraphics[width=\columnwidth]{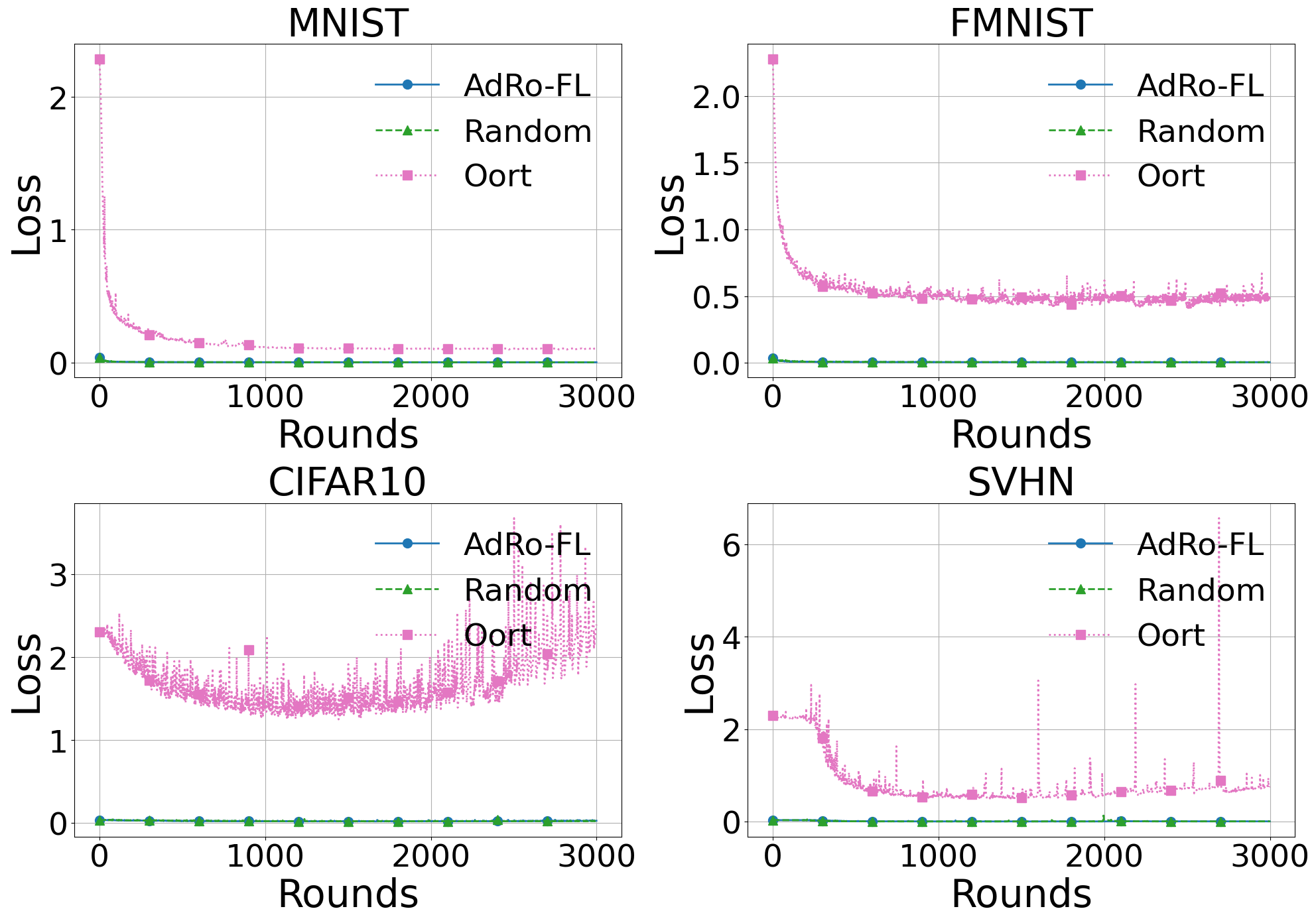}
    \caption{Loss comparison of AdRo-FL with random selection and Oort on four datasets. Here we used global selection (cluster-oriented).}
    \label{loss_comp_global_cluster}
\end{figure}

While global selection in AdRo-FL, although the aggregator ranks clients globally, each cluster-head enforces the privacy constraint locally by checking if at least $C$ clients from its cluster were selected. If not, it withholds participation, thereby ensuring privacy. Fig. ~\ref{acc_comp_global_cluster} shows the accuracy of AdRo-FL and the baselines against the FL round numbers. The figures shows that AdRo-FL either achieves better or competitive performance compared to baselines. Fig. ~\ref{loss_comp_global_cluster} shows the loss values of AdRo-FL and the baselines against FL round numbers.

\begin{table}[H]
\caption{Time-to-accuracy and final test accuracy comparison on the MNIST dataset (Cluster-oriented setting). The best value for each row is highlighted in bold. Here clients were selected globally from all clusters.}
\label{tab:MNIST_global_cluster_results}
\centering
\resizebox{\columnwidth}{!}{%
\begin{tabular}{lcccccccc}
\hline
\textbf{Method} & \textbf{@ 60} & \textbf{@ 65} & \textbf{@ 70} & \textbf{@ 75} & \textbf{@ 80} & \textbf{@ 85} & \textbf{@ 90} & \textbf{Best Accuracy (\%)} \\
\hline
AdRo-FL & \textbf{15} & \textbf{15} & \textbf{15} & 23 & 37 & 45 & \textbf{74} & 98.19 \\
Random   & 21 & 21 & 21 & \textbf{25} & \textbf{36} & \textbf{45} & 77 & \textbf{98.23} \\
Oort     & 18 & 19 & 28 & 28 & 38 & 47 & 108 & 97.25 \\
\hline
\end{tabular}%
}
\end{table}

In table \ref{tab:MNIST_global_cluster_results}, we show the number of rounds needed to reach different target accuracies for MNIST dataset. Also we note the final accuracy achieved by each method (proposed and Oort). On average, compared to random baseline, AdRo-FL achieves $1.19\times$ time-to-accuracy improvement. Compared to Oort, it achieves $1.30\times$ time-to-accuracy improvement on average and $1.01\times$ final accuracy improvement.

\begin{table}[H]
\caption{Time-to-accuracy and final test accuracy comparison on the FMNIST dataset (Cluster-oriented setting). The best value for each row is highlighted in bold. Here clients were selected globally from all clusters.}
\label{tab:FMNIST_global_cluster_results}
\centering
\resizebox{\columnwidth}{!}{%
\begin{tabular}{lccccccc}
\hline
\textbf{Method} & \textbf{@ 60} & \textbf{@ 65} & \textbf{@ 70} & \textbf{@ 75} & \textbf{@ 80} & \textbf{@ 85} & \textbf{Best Accuracy (\%)} \\
\hline
AdRo-FL & 40 & \textbf{40} & 67 & \textbf{118} & \textbf{243} & \textbf{795} & \textbf{88.06} \\
Random   & 50 & 87 & 122 & 167 & 312 & 980 & 87.47 \\
Oort     & \textbf{37} & 58 & \textbf{78} & 157 & 348 & 2195 & 85.82 \\
\hline
\end{tabular}%
}
\end{table}

In table \ref{tab:FMNIST_global_cluster_results}, we show the number of rounds needed to reach different target accuracies for FMNIST dataset. Also we note the final accuracy achieved by each method (proposed and Oort). On average, compared to random baseline, AdRo-FL achieves $1.53\times$ time-to-accuracy improvement and $1.01\times$ final accuracy improvement. Compared to Oort, it achieves $1.51\times$ time-to-accuracy improvement on average and $1.03\times$ final accuracy improvement.

\begin{table}[H]
\caption{Time-to-accuracy and final test accuracy comparison on the CIFAR10 dataset (Cluster-oriented setting). The best value for each row is highlighted in bold. Here clients were selected globally from all clusters.}
\label{tab:CIFAR10_global_cluster_results}
\centering
\resizebox{\columnwidth}{!}{%
\begin{tabular}{lcccccc}
\hline
\textbf{Method} & \textbf{@ 35} & \textbf{@ 40} & \textbf{@ 45} & \textbf{@ 50} & \textbf{@ 55} & \textbf{Best Accuracy (\%)} \\
\hline
AdRo-FL & 283 & 347 & 496 & 728 & 1108 & \textbf{58.68} \\
Random   & 408 & 497 & 666 & 974 & 1557 & 57.90 \\
Oort     & \textbf{210} & \textbf{313} & \textbf{397} & \textbf{655} & \textbf{988} & 56.69 \\
\hline
\end{tabular}%
}
\end{table}

In table \ref{tab:CIFAR10_global_cluster_results}, we show the number of rounds needed to reach different target accuracies for CIFAR10 dataset. Also we note the final accuracy achieved by each method (proposed and Oort). On average, compared to random baseline, AdRo-FL achieves $1.39\times$ time-to-accuracy improvement and $1.01\times$ final accuracy improvement. Compared to Oort, it achieves $1.04\times$ final accuracy improvement.

\begin{table}[H]
\caption{Time-to-accuracy and final test accuracy comparison on the SVHN dataset (Cluster-oriented setting). The best value for each row is highlighted in bold. Here clients were selected globally from all clusters.}
\label{tab:SVHN_global_cluster_results}
\centering
\resizebox{\columnwidth}{!}{%
\begin{tabular}{lccccccc}
\hline
\textbf{Method} & \textbf{@ 60} & \textbf{@ 65} & \textbf{@ 70} & \textbf{@ 75} & \textbf{@ 80} & \textbf{@ 85} & \textbf{Best Accuracy (\%)} \\
\hline
AdRo-FL & 344 & 358 & 387 & 418 & 499 & 698 & 88.46 \\
Random   & \textbf{285} & \textbf{300} & \textbf{335} & \textbf{363} & \textbf{441} & \textbf{627} & \textbf{88.51} \\
Oort     & 335 & 351 & 376 & 414 & 512 & 873 & 86.84 \\
\hline
\end{tabular}%
}
\end{table}

In table \ref{tab:SVHN_global_cluster_results}, we show the number of rounds needed to reach different target accuracies for SVHN dataset. Also we note the final accuracy achieved by each method (proposed and Oort). Compared to Oort, it achieves $1.03\times$ time-to-accuracy improvement on average and $1.02\times$ final accuracy improvement.

\begin{figure}[H]
    \centering
    \includegraphics[width=\columnwidth]{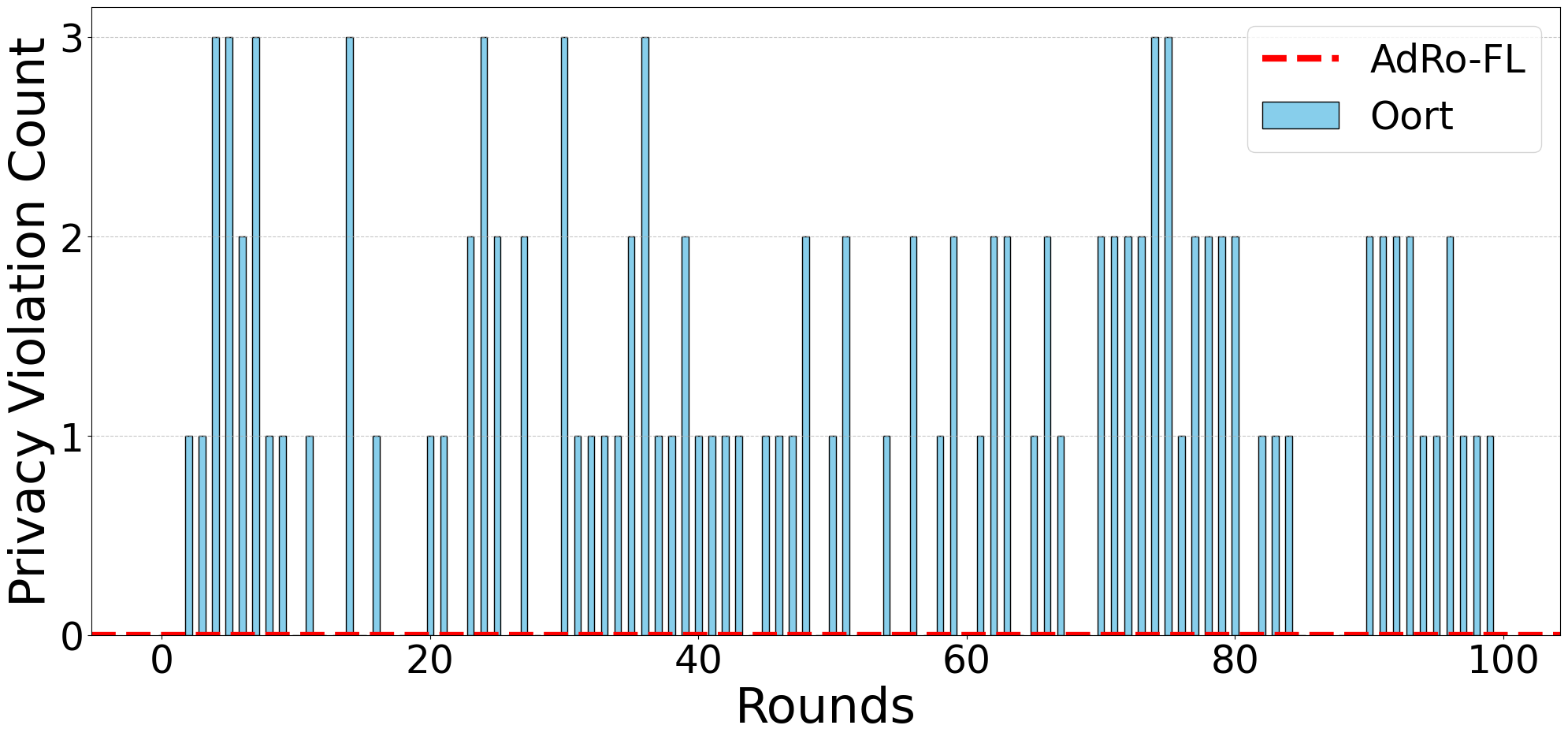}
    \caption{Number of clusters became vulnerable to BSA per round.}
    \label{privacy_violation_count}
\end{figure}

Lastly, in Fig. \ref{privacy_violation_count}, we show how many clusters became vulnerable to BSA per round because of violating cluster constrains when doing client selection globally in the cluster-oriented setting. We discussed in section \ref{threatmodel} and \ref{prevent_bsa_in_cluster} that if the adversarial aggregator can manage to single out a client from a cluster, it can do BSA to know clients private model update. Insecure client selection technique Oort does not consider this important constraint while selecting best clients based on utility. The figure illustrates the privacy risk of Oort if applied on cluster-oriented client setting to select client globally. Compared to Oort, AdRo-FL handles this vulnerability while not compromising on model performance.

\subsubsection{Distributed, Non-cluster-oriented Client Setting}
\begin{figure}[H]
    \centering
    \includegraphics[width=\columnwidth]{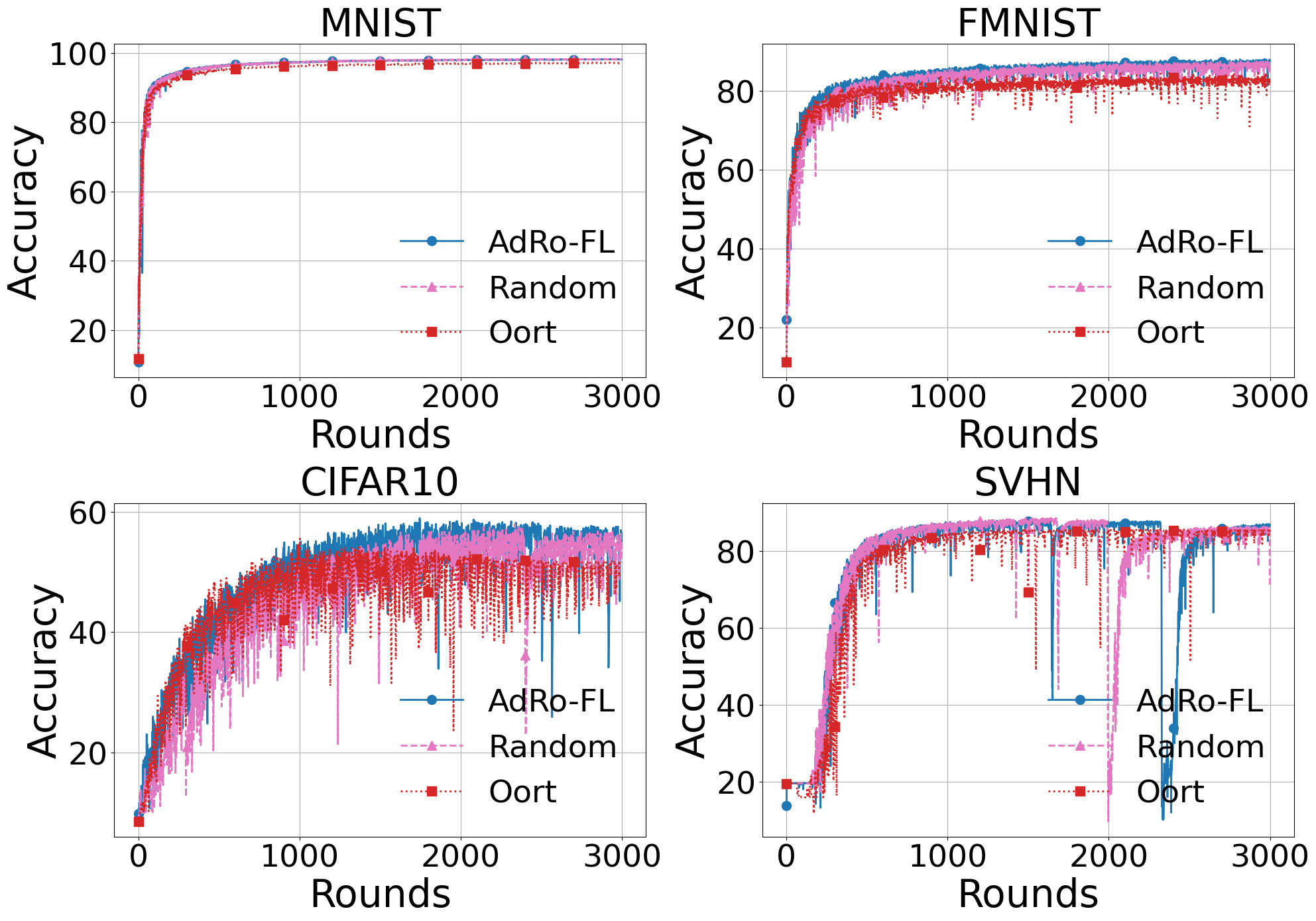}
    \caption{Accuracy comparison of AdRo-FL (VRF-based) with random selection and Oort on four datasets in non-cluster-oriented setting.}
    \label{acc_comp_vrf}
\end{figure}

\begin{figure}[H]
    \centering
    \includegraphics[width=\columnwidth]{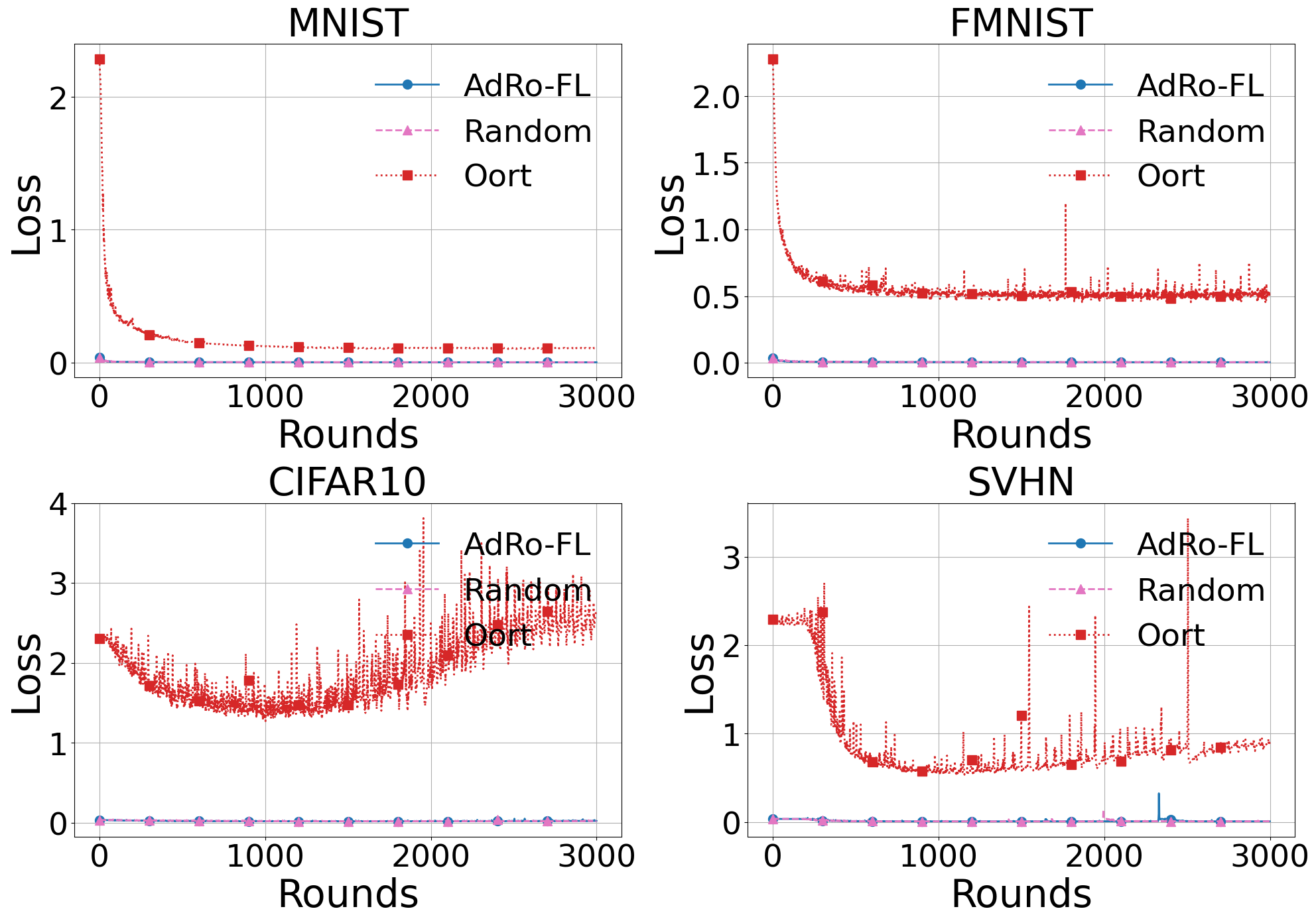}
    \caption{Loss comparison of AdRo-FL (VRF-based) with random selection and Oort on four datasets in non-cluster-oriented setting.}
    \label{loss_comp_vrf}
\end{figure}

In this section, we present the results of experiments conducted with non-cluster-oriented clients. As detailed in section \ref{informed_selection_VRF}, AdRo-FL uses two level client filtering for informed client selection in distributed, non-cluster-oriented client setting. We compared the result with baselines. In Fig. \ref{acc_comp_vrf}, we show result of 3000 rounds of training on the selected datasets. The plot shows that AdRo-FL achieves better time-to-accuracy as well as target accuracy than Oort. In Fig. \ref{loss_comp_vrf}, we shows the loss reduction across FL rounds.

\begin{table}[H]
\caption{Time-to-accuracy and final test accuracy comparison on the MNIST dataset (non-cluster-oriented setting). Best values for each row are highlighted in bold.}
\centering
\resizebox{\columnwidth}{!}{%
\begin{tabular}{lrrrrrrrl}
\hline
\textbf{Method} & @{\,}\textbf{60} & @{\,}\textbf{65} & @{\,}\textbf{70} & @{\,}\textbf{75} & @{\,}\textbf{80} & @{\,}\textbf{85} & @{\,}\textbf{90} & \textbf{Best Accuracy (\%)} \\
\hline
AdRo-FL & \textbf{15} & \textbf{15} & \textbf{15} & \textbf{23} & 37 & \textbf{45} & \textbf{74} & 98.19 \\
Random   & 21 & 21 & 21 & 25 & \textbf{36} & \textbf{45} & 77 & \textbf{98.23} \\
Oort     & 16 & 21 & 22 & 25 & \textbf{36} & 52 & 112 & 97.17 \\
\hline
\end{tabular}%
}
\label{tab:MNIST_noncluster_results}
\end{table}

In table \ref{tab:MNIST_noncluster_results}, we show the number of rounds needed to reach different target accuracies for MNIST dataset. Also we note the final accuracy achieved by each method. On average, compared to Oort, AdRo-FL achieves $1.24\times$ time-to-accuracy improvement and $1.01\times$ final accuracy improvement. Compared to random selection, AdRo-FL achieves $1.19\times$ time-to-accuracy improvement on average.

\begin{table}[H]
\caption{Time-to-accuracy and final test accuracy comparison on the FMNIST dataset (non-cluster-oriented setting). Best values for each row are highlighted in bold.}
\centering
\resizebox{\columnwidth}{!}{%
\begin{tabular}{lrrrrrrl}
\hline
\textbf{Method} & @{\,}\textbf{60} & @{\,}\textbf{65} & @{\,}\textbf{70} & @{\,}\textbf{75} & @{\,}\textbf{80} & @{\,}\textbf{85} & \textbf{Best Accuracy (\%)} \\
\hline
AdRo-FL & 40 & \textbf{40} & \textbf{67} & \textbf{118} & \textbf{243} & \textbf{795} & \textbf{88.06} \\
Random   & 50 & 87 & 122 & 167 & 312 & 980 & 87.47 \\
Oort     & \textbf{36} & 55 & 93 & 128 & 345 & N/A & 84.59 \\
\hline
\end{tabular}%
}
\label{tab:FMNIST_noncluster_results}
\end{table}

In table \ref{tab:FMNIST_noncluster_results}, we show the number of rounds needed to reach different target accuracies for FMNIST dataset. Also we note the final accuracy achieved by each method. On average, compared to Oort, AdRo-FL achieves $1.23\times$ time-to-accuracy improvement and $1.04\times$ final accuracy improvement. Compared to random selection, AdRo-FL achieves $1.53\times$ time-to-accuracy improvement and $1.01\times$ final accuracy improvement, on average.

\begin{table}[H]
\caption{Time-to-accuracy and final test accuracy comparison on the CIFAR-10 dataset (non-cluster-oriented setting). Best values for each row are highlighted in bold.}
\centering
\resizebox{\columnwidth}{!}{%
\begin{tabular}{lrrrrrl}
\hline
\textbf{Method} & @{\,}\textbf{35} & @{\,}\textbf{40} & @{\,}\textbf{45} & @{\,}\textbf{50} & @{\,}\textbf{55} & \textbf{Best Accuracy (\%)} \\
\hline
AdRo-FL & \textbf{219} & 336 & \textbf{415} & 739 & \textbf{966} & \textbf{58.92} \\
Random   & 408 & 497 & 666 & 974 & 1557 & 57.90 \\
Oort     & 225 & \textbf{286} & 428 & \textbf{674} & 1003 & 55.53 \\
\hline
\end{tabular}%
}
\label{tab:CIFAR10_noncluster_results}
\end{table}

In table \ref{tab:CIFAR10_noncluster_results}, we show the number of rounds needed to reach different target accuracies for CIFAR10 dataset. Also we note the final accuracy achieved by each method. On average, compared to Oort, AdRo-FL achieves $1.06\times$ final accuracy improvement. However, compared to random selection, on average, AdRo-FL achieves $1.58\times$ time-to-accuracy and $1.02\times$ final accuracy improvement.

\begin{table}[H]
\caption{Time-to-accuracy and final test accuracy comparison on the SVHN dataset (non-cluster-oriented setting). Best values for each row are highlighted in bold.}
\centering
\resizebox{\columnwidth}{!}{%
\begin{tabular}{lrrrrrrl}
\hline
\textbf{Method} & @{\,}\textbf{60} & @{\,}\textbf{65} & @{\,}\textbf{70} & @{\,}\textbf{75} & @{\,}\textbf{80} & @{\,}\textbf{85} & \textbf{Best Accuracy (\%)} \\
\hline
AdRo-FL & \textbf{272} & 301 & \textbf{326} & \textbf{361} & \textbf{422} & 643 & 87.99 \\
Random   & 285 & \textbf{300} & 335 & 363 & 441 & \textbf{627} & \textbf{88.51} \\
Oort     & 347 & 355 & 397 & 439 & 552 & 1158 & 85.53 \\
\hline
\end{tabular}%
}
\label{tab:SVHN_global_results}
\end{table}

In table \ref{tab:SVHN_global_results}, we show the number of rounds needed to reach different target accuracies for SVHN dataset. Also we note the final accuracy achieved by each method. On average, compared to Oort, AdRo-FL achieves $1.33\times$ time-to-accuracy improvement and $1.03\times$ final accuracy improvement. Compared to random selection, time-to-accuracy was improved $1.02\times$ on average.

%As for the time needed for VRF hash and proof generation in our experiments is 0.24 seconds on average per client. The VRF proof verification time for one client is 0.18 seconds on average. This could be improved by using faster VRF implementation. For experimental purpose, we used a simple implementation~\cite{Nccgroup} of VRF. According to the authors of the VRF implementation, `` This code is alpha-quality and is not suitable for production...Specifically, both the algorithms within the code and (the use of) Python's big integers are clearly not constant time and thus introduce timing side channels." As for the client utility signature generation time (needed in the first level of client selection for non-cluster-oriented clients) for one client, it was 0.000015 seconds on average. The signature verification time for one utility value was 0.000064 seconds on average including the decompression time of the whole signature string received from the aggregator.

\subsubsection{Resource Consumption Comparison}
\begin{figure}[H]
    \centering
    \includegraphics[width=\columnwidth]{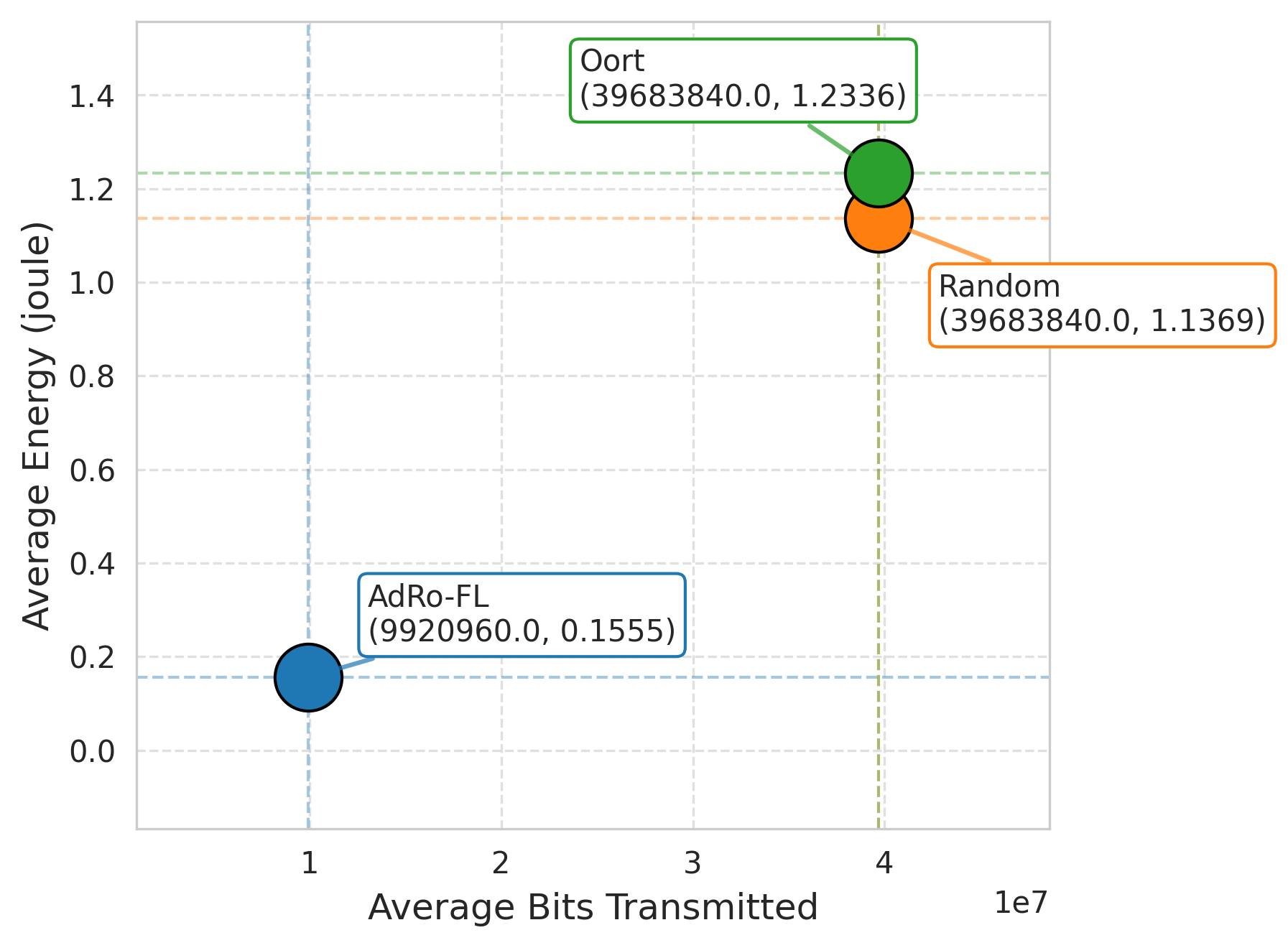}
    \caption{Average energy consumption (Jules) and bits transmitted per round for SVHN dataset in cluster-oriented setting.}
    \label{energy_bits_comp}
\end{figure}

In Fig. \ref{energy_bits_comp}, we showed the energy consumption estimations. The figure shows energy consumption estimated in cluster-oriented client setting. However, similar results were observed for non-cluster-oriented client setting. AdRo-FL achieved superior energy efficiency due to deadline based client selection and quantization. This energy efficiency is achieved without noticeable degradation of performance which indicates AdRo-FL's robustness in delivering better performance with reduced memory consumption.

\section{Conclusion}
\label{C}
Ensuring faster convergence as well as client privacy remains one of the most pressing challenges in FL, particularly in adversarial settings where the central aggregator may not follow the protocol. Ensuring the aggregator cannot alter the client selection process to do biased selection attack, is still a critical challenge that requires robust mechanisms to guarantee fairness and verifiability in FL systems. In this work we presented AdRo-FL, a practical solution to an underexplored problem of secure and informed client selection in the presence of adversarial aggregator in FL. Unlike existing methods, which either compromise performance through random uninformed client selection or risk privacy by relying on utility-based informed selection, we uniquely cover both that realizes informed as well as secure client selection. AdRo-FL prevents adversarial aggregator from singling out a victim client while ensuring that clients are selected based on their practical utility. Experimental outcomes with several benchmark datasets shows that AdRo-FL securely achieves significant performance improvement over insecure baselines.

% \section*{Funding Acknowledgment}
% This research is supported by a grant (No. CRPG-00-0000) under the Cybersecurity Research and Innovation Pioneers Initiative, provided by the National Cybersecurity Authority (NCA) in the Kingdom of Saudi Arabia.

%\vspace{12pt}
\bibliographystyle{IEEEtran}
\bibliography{references}

\begin{thebibliography}{10}
\providecommand{\url}[1]{#1}
\csname url@samestyle\endcsname
\providecommand{\newblock}{\relax}
\providecommand{\bibinfo}[2]{#2}
\providecommand{\BIBentrySTDinterwordspacing}{\spaceskip=0pt\relax}
\providecommand{\BIBentryALTinterwordstretchfactor}{4}
\providecommand{\BIBentryALTinterwordspacing}{\spaceskip=\fontdimen2\font plus
\BIBentryALTinterwordstretchfactor\fontdimen3\font minus \fontdimen4\font\relax}
\providecommand{\BIBforeignlanguage}[2]{{%
\expandafter\ifx\csname l@#1\endcsname\relax
\typeout{** WARNING: IEEEtran.bst: No hyphenation pattern has been}%
\typeout{** loaded for the language `#1'. Using the pattern for}%
\typeout{** the default language instead.}%
\else
\language=\csname l@#1\endcsname
\fi
#2}}
\providecommand{\BIBdecl}{\relax}
\BIBdecl

\bibitem{zhang2021survey}
C.~Zhang, Y.~Xie, H.~Bai, B.~Yu, W.~Li, and Y.~Gao, ``A survey on federated learning,'' \emph{Knowledge-Based Systems}, vol. 216, p. 106775, 2021.

\bibitem{le2023privacy}
J.~Le, D.~Zhang, X.~Lei, L.~Jiao, K.~Zeng, and X.~Liao, ``Privacy-preserving federated learning with malicious clients and honest-but-curious servers,'' \emph{IEEE Transactions on Information Forensics and Security}, 2023.

\bibitem{CAI2024102420}
\BIBentryALTinterwordspacing
J.~Cai, W.~Shen, and J.~Qin, ``Esvfl: Efficient and secure verifiable federated learning with privacy-preserving,'' \emph{Information Fusion}, vol. 109, p. 102420, 2024. [Online]. Available: \url{https://www.sciencedirect.com/science/article/pii/S1566253524001982}
\BIBentrySTDinterwordspacing

\bibitem{9464278}
P.~Kairouz, H.~B. McMahan, B.~Avent, A.~Bellet, M.~Bennis, A.~N. Bhagoji, K.~Bonawit, Z.~Charles, G.~Cormode, R.~Cummings, R.~G.~L. D’Oliveira, H.~Eichner, S.~El~Rouayheb, D.~Evans, J.~Gardner, Z.~Garrett, A.~Gascón, B.~Ghazi, P.~B. Gibbons, M.~Gruteser, Z.~Harchaoui, C.~He, L.~He, Z.~Huo, B.~Hutchinson, J.~Hsu, M.~Jaggi, T.~Javidi, G.~Joshi, M.~Khodak, J.~Konecný, A.~Korolova, F.~Koushanfar, S.~Koyejo, T.~Lepoint, Y.~Liu, P.~Mittal, M.~Mohri, R.~Nock, A.~Özgür, R.~Pagh, H.~Qi, D.~Ramage, R.~Raskar, M.~Raykova, D.~Song, W.~Song, S.~U. Stich, Z.~Sun, A.~Theertha~Suresh, F.~Tramèr, P.~Vepakomma, J.~Wang, L.~Xiong, Z.~Xu, Q.~Yang, F.~X. Yu, H.~Yu, and S.~Zhao, 2021.

\bibitem{fi15090310}
\BIBentryALTinterwordspacing
R.~Aziz, S.~Banerjee, S.~Bouzefrane, and T.~Le~Vinh, ``Exploring homomorphic encryption and differential privacy techniques towards secure federated learning paradigm,'' \emph{Future Internet}, vol.~15, no.~9, 2023. [Online]. Available: \url{https://www.mdpi.com/1999-5903/15/9/310}
\BIBentrySTDinterwordspacing

\bibitem{RODRIGUEZBARROSO2020270}
\BIBentryALTinterwordspacing
N.~Rodríguez-Barroso, G.~Stipcich, D.~Jiménez-López, J.~A. Ruiz-Millán, E.~Martínez-Cámara, G.~González-Seco, M.~V. Luzón, M.~A. Veganzones, and F.~Herrera, ``Federated learning and differential privacy: Software tools analysis, the sherpa.ai fl framework and methodological guidelines for preserving data privacy,'' \emph{Information Fusion}, vol.~64, pp. 270--292, 2020. [Online]. Available: \url{https://www.sciencedirect.com/science/article/pii/S1566253520303213}
\BIBentrySTDinterwordspacing

\bibitem{10.1145/3133956.3133982}
\BIBentryALTinterwordspacing
K.~Bonawitz, V.~Ivanov, B.~Kreuter, A.~Marcedone, H.~B. McMahan, S.~Patel, D.~Ramage, A.~Segal, and K.~Seth, ``Practical secure aggregation for privacy-preserving machine learning,'' in \emph{Proceedings of the 2017 ACM SIGSAC Conference on Computer and Communications Security}, ser. CCS '17.\hskip 1em plus 0.5em minus 0.4em\relax New York, NY, USA: Association for Computing Machinery, 2017, p. 1175–1191. [Online]. Available: \url{https://doi.org/10.1145/3133956.3133982}
\BIBentrySTDinterwordspacing

\bibitem{nguyen2022blockchain}
T.~Nguyen, P.~Thai, J.~Tre’R, T.~N. Dinh, and M.~T. Thai, ``Blockchain-based secure client selection in federated learning,'' in \emph{2022 IEEE International Conference on Blockchain and Cryptocurrency (ICBC)}.\hskip 1em plus 0.5em minus 0.4em\relax IEEE, 2022, pp. 1--9.

\bibitem{298104}
\BIBentryALTinterwordspacing
Z.~Jiang, P.~Ye, S.~He, W.~Wang, R.~Chen, and B.~Li, ``Lotto: Secure participant selection against adversarial servers in federated learning,'' in \emph{33rd USENIX Security Symposium (USENIX Security 24)}.\hskip 1em plus 0.5em minus 0.4em\relax Philadelphia, PA: USENIX Association, Aug. 2024, pp. 343--360. [Online]. Available: \url{https://www.usenix.org/conference/usenixsecurity24/presentation/jiang-zhifeng}
\BIBentrySTDinterwordspacing

\bibitem{li2024comprehensive}
J.~Li, T.~Chen, and S.~Teng, ``A comprehensive survey on client selection strategies in federated learning,'' \emph{Computer Networks}, p. 110663, 2024.

\bibitem{jimenez2024fedartml}
G.~D.~M. Jimenez, A.~Anagnostopoulos, I.~Chatzigiannakis, and A.~Vitaletti, ``Fedartml: A tool to facilitate the generation of non-iid datasets in a controlled way to support federated learning research,'' \emph{IEEE Access}, 2024.

\bibitem{273723}
\BIBentryALTinterwordspacing
F.~Lai, X.~Zhu, H.~V. Madhyastha, and M.~Chowdhury, ``Oort: Efficient federated learning via guided participant selection,'' in \emph{15th {USENIX} Symposium on Operating Systems Design and Implementation ({OSDI} 21)}.\hskip 1em plus 0.5em minus 0.4em\relax {USENIX} Association, Jul. 2021, pp. 19--35. [Online]. Available: \url{https://www.usenix.org/conference/osdi21/presentation/lai}
\BIBentrySTDinterwordspacing

\bibitem{li2024node}
Z.~Li, Y.~Dang, and X.~Chen, ``Node selection for model quality optimization in hierarchical federated learning based on deep reinforcement learning,'' \emph{Peer-to-Peer Networking and Applications}, vol.~17, no.~3, pp. 1720--1731, 2024.

\bibitem{tang2021fedgp}
M.~Tang, X.~Ning, Y.~Wang, Y.~Wang, and Y.~Chen, ``Fedgp: Correlation-based active client selection strategy for heterogeneous federated learning,'' \emph{arXiv preprint arXiv:2103.13822}, 2021.

\bibitem{wu2022node}
H.~Wu and P.~Wang, ``Node selection toward faster convergence for federated learning on non-iid data,'' \emph{IEEE Transactions on Network Science and Engineering}, vol.~9, no.~5, pp. 3099--3111, 2022.

\bibitem{10534777}
S.~Trindade and N.~L.~S. da~Fonseca, ``Client selection in hierarchical federated learning,'' \emph{IEEE Internet of Things Journal}, vol.~11, no.~17, pp. 28\,480--28\,495, 2024.

\bibitem{9846900}
F.~Shi, C.~Hu, W.~Lin, L.~Fan, T.~Huang, and W.~Wu, ``Vfedcs: Optimizing client selection for volatile federated learning,'' \emph{IEEE Internet of Things Journal}, vol.~9, no.~24, pp. 24\,995--25\,010, 2022.

\bibitem{10589575}
X.~Chen, X.~Zhou, H.~Zhang, M.~Sun, and H.~Vincent~Poor, ``Client selection for wireless federated learning with data and latency heterogeneity,'' \emph{IEEE Internet of Things Journal}, vol.~11, no.~19, pp. 32\,183--32\,196, 2024.

\bibitem{10145997}
O.~Wehbi, S.~Arisdakessian, O.~A. Wahab, H.~Otrok, S.~Otoum, A.~Mourad, and M.~Guizani, ``Fedmint: Intelligent bilateral client selection in federated learning with newcomer iot devices,'' \emph{IEEE Internet of Things Journal}, vol.~10, no.~23, pp. 20\,884--20\,898, 2023.

\bibitem{marnissi2024client}
O.~Marnissi, H.~E. Hammouti, and E.~H. Bergou, ``Client selection in federated learning based on gradients importance,'' in \emph{AIP Conference Proceedings}, vol. 3034, no.~1.\hskip 1em plus 0.5em minus 0.4em\relax AIP Publishing, 2024.

\bibitem{cho2022towards}
Y.~J. Cho, J.~Wang, and G.~Joshi, ``Towards understanding biased client selection in federated learning,'' in \emph{International Conference on Artificial Intelligence and Statistics}.\hskip 1em plus 0.5em minus 0.4em\relax PMLR, 2022, pp. 10\,351--10\,375.

\bibitem{SABAH2025102756}
\BIBentryALTinterwordspacing
F.~Sabah, Y.~Chen, Z.~Yang, A.~Raheem, M.~Azam, N.~Ahmad, and R.~Sarwar, ``Fairdpfl-scs: Fair dynamic personalized federated learning with strategic client selection for improved accuracy and fairness,'' \emph{Information Fusion}, vol. 115, p. 102756, 2025. [Online]. Available: \url{https://www.sciencedirect.com/science/article/pii/S1566253524005347}
\BIBentrySTDinterwordspacing

\bibitem{GUO2024102549}
\BIBentryALTinterwordspacing
J.~Guo, L.~Su, J.~Liu, J.~Ding, X.~Liu, B.~Huang, and L.~Li, ``Auction-based client selection for online federated learning,'' \emph{Information Fusion}, vol. 112, p. 102549, 2024. [Online]. Available: \url{https://www.sciencedirect.com/science/article/pii/S1566253524003270}
\BIBentrySTDinterwordspacing

\bibitem{10643330}
Z.~Niu, H.~Dong, A.~K. Qin, and T.~Gu, ``Flrce: Resource-efficient federated learning with early-stopping strategy,'' \emph{IEEE Transactions on Mobile Computing}, pp. 1--16, 2024.

\bibitem{DBLP:journals/corr/Dettmers15}
T.~Dettmers, ``8-bit approximations for parallelism in deep learning,'' in \emph{4th International Conference on Learning Representations, {ICLR}, San Juan, Puerto Rico}, Y.~Bengio and Y.~LeCun, Eds., 2016.

\bibitem{li2022pyramidfl}
C.~Li, X.~Zeng, M.~Zhang, and Z.~Cao, ``Pyramidfl: A fine-grained client selection framework for efficient federated learning,'' in \emph{Proceedings of the 28th Annual International Conference on Mobile Computing And Networking}, 2022, pp. 158--171.

\bibitem{briggs2020federated}
C.~Briggs, Z.~Fan, and P.~Andras, ``Federated learning with hierarchical clustering of local updates to improve training on non-iid data,'' in \emph{2020 international joint conference on neural networks (IJCNN)}.\hskip 1em plus 0.5em minus 0.4em\relax IEEE, 2020, pp. 1--9.

\bibitem{9610118}
M.~Asad, A.~Moustafa, F.~A. Rabhi, and M.~Aslam, ``Thf: 3-way hierarchical framework for efficient client selection and resource management in federated learning,'' \emph{IEEE Internet of Things Journal}, vol.~9, no.~13, pp. 11\,085--11\,097, 2022.

\bibitem{10.1145/3620678.3624651}
\BIBentryALTinterwordspacing
J.~Liu, F.~Lai, Y.~Dai, A.~Akella, H.~V. Madhyastha, and M.~Chowdhury, ``Auxo: Efficient federated learning via scalable client clustering,'' in \emph{Proceedings of the 2023 ACM Symposium on Cloud Computing}, ser. SoCC '23.\hskip 1em plus 0.5em minus 0.4em\relax New York, NY, USA: Association for Computing Machinery, 2023, p. 125–141. [Online]. Available: \url{https://doi.org/10.1145/3620678.3624651}
\BIBentrySTDinterwordspacing

\bibitem{10083200}
C.~Keçeci, M.~Shaqfeh, F.~Al-Qahtani, M.~Ismail, and E.~Serpedin, ``Clustered scheduling and communication pipelining for efficient resource management of wireless federated learning,'' \emph{IEEE Internet of Things Journal}, vol.~10, no.~15, pp. 13\,303--13\,316, 2023.

\bibitem{MLSYS2022_6c44dc73}
\BIBentryALTinterwordspacing
J.~So, C.~He, C.-S. Yang, S.~Li, Q.~Yu, R.~E.~Ali, B.~Guler, and S.~Avestimehr, ``Lightsecagg: a lightweight and versatile design for secure aggregation in federated learning,'' in \emph{Proceedings of Machine Learning and Systems}, D.~Marculescu, Y.~Chi, and C.~Wu, Eds., vol.~4, 2022, pp. 694--720. [Online]. Available: \url{https://proceedings.mlsys.org/paper_files/paper/2022/file/6c44dc73014d66ba49b28d483a8f8b0d-Paper.pdf}
\BIBentrySTDinterwordspacing

\bibitem{10.5555/3310435.3310586}
U.~Erlingsson, V.~Feldman, I.~Mironov, A.~Raghunathan, K.~Talwar, and A.~Thakurta, ``Amplification by shuffling: from local to central differential privacy via anonymity,'' in \emph{Proceedings of the Thirtieth Annual ACM-SIAM Symposium on Discrete Algorithms}, ser. SODA '19.\hskip 1em plus 0.5em minus 0.4em\relax USA: Society for Industrial and Applied Mathematics, 2019, p. 2468–2479.

\bibitem{10.1145/3627703.3629559}
\BIBentryALTinterwordspacing
Z.~Jiang, W.~Wang, and R.~Chen, ``Dordis: Efficient federated learning with dropout-resilient differential privacy,'' ser. EuroSys '24.\hskip 1em plus 0.5em minus 0.4em\relax New York, NY, USA: Association for Computing Machinery, 2024, p. 472–488. [Online]. Available: \url{https://doi.org/10.1145/3627703.3629559}
\BIBentrySTDinterwordspacing

\bibitem{fung2020limitations}
C.~Fung, C.~J. Yoon, and I.~Beschastnikh, ``The limitations of federated learning in sybil settings,'' in \emph{23rd International Symposium on Research in Attacks, Intrusions and Defenses (RAID 2020)}, 2020, pp. 301--316.

\bibitem{10.1007/978-3-030-60548-3_18}
\BIBentryALTinterwordspacing
H.~R. Roth, K.~Chang, P.~Singh, N.~Neumark, W.~Li, V.~Gupta, S.~Gupta, L.~Qu, A.~Ihsani, B.~C. Bizzo, Y.~Wen, V.~Buch, M.~Shah, F.~Kitamura, M.~Mendon\c{c}a, V.~Lavor, A.~Harouni, C.~Compas, J.~Tetreault, P.~Dogra, Y.~Cheng, S.~Erdal, R.~White, B.~Hashemian, T.~Schultz, M.~Zhang, A.~McCarthy, B.~M. Yun, E.~Sharaf, K.~V. Hoebel, J.~B. Patel, B.~Chen, S.~Ko, E.~Leibovitz, E.~D. Pisano, L.~Coombs, D.~Xu, K.~J. Dreyer, I.~Dayan, R.~C. Naidu, M.~Flores, D.~Rubin, and J.~Kalpathy-Cramer, ``Federated learning for breast density classification: A real-world implementation.''\hskip 1em plus 0.5em minus 0.4em\relax Berlin, Heidelberg: Springer-Verlag, 2020, p. 181–191. [Online]. Available: \url{https://doi.org/10.1007/978-3-030-60548-3_18}
\BIBentrySTDinterwordspacing

\bibitem{said2023hipaa}
A.~Said, A.~Yahyaoui, and T.~Abdellatif, ``Hipaa and gdpr compliance in iot healthcare systems,'' in \emph{International conference on model and data engineering}.\hskip 1em plus 0.5em minus 0.4em\relax Springer, 2023, pp. 198--209.

\bibitem{brendel2021provable}
J.~Brendel, C.~Cremers, D.~Jackson, and M.~Zhao, ``The provable security of ed25519: theory and practice,'' in \emph{2021 IEEE Symposium on Security and Privacy (SP)}.\hskip 1em plus 0.5em minus 0.4em\relax IEEE, 2021, pp. 1659--1676.

\bibitem{maulidina2023comparative}
A.~P. Maulidina, R.~A. Wijaya, K.~Mazel, and M.~S. Astriani, ``Comparative study of data compression algorithms: Zstandard, zlib \& lz4,'' in \emph{International Conference on Science, Engineering Management and Information Technology}.\hskip 1em plus 0.5em minus 0.4em\relax Springer, 2023, pp. 394--406.

\bibitem{papadopoulos2017making}
D.~Papadopoulos, D.~Wessels, S.~Huque, M.~Naor, J.~V{\v{c}}el{\'a}k, L.~Reyzin, and S.~Goldberg, ``Making nsec5 practical for dnssec,'' \emph{Cryptology ePrint Archive}, 2017.

\bibitem{mcmahan2017communication}
B.~McMahan, E.~Moore, D.~Ramage, S.~Hampson, and B.~A. y~Arcas, ``Communication-efficient learning of deep networks from decentralized data,'' in \emph{Artificial intelligence and statistics}.\hskip 1em plus 0.5em minus 0.4em\relax PMLR, 2017, pp. 1273--1282.

\bibitem{10.5555/3666122.3669518}
Z.~Xiao, Z.~Chen, S.~Liu, H.~Wang, Y.~Feng, J.~Hao, J.~T. Zhou, J.~Wu, H.~H. Yang, and Z.~Liu, ``Fed-grab: federated long-tailed learning with self-adjusting gradient balancer,'' in \emph{Proceedings of the 37th International Conference on Neural Information Processing Systems}, ser. NIPS '23.\hskip 1em plus 0.5em minus 0.4em\relax Red Hook, NY, USA: Curran Associates Inc., 2023.

\bibitem{9833969}
M.~Mortaheb, C.~Vahapoglu, and S.~Ulukus, ``Fedgradnorm: Personalized federated gradient-normalized multi-task learning,'' in \emph{2022 IEEE 23rd International Workshop on Signal Processing Advances in Wireless Communication (SPAWC)}, 2022, pp. 1--5.

\bibitem{FEKRI2023109285}
\BIBentryALTinterwordspacing
M.~N. Fekri, K.~Grolinger, and S.~Mir, ``Asynchronous adaptive federated learning for distributed load forecasting with smart meter data,'' \emph{International Journal of Electrical Power \& Energy Systems}, vol. 153, p. 109285, 2023. [Online]. Available: \url{https://www.sciencedirect.com/science/article/pii/S0142061523003423}
\BIBentrySTDinterwordspacing

\end{thebibliography}

\appendices
\section{Verifiable Random Function (VRF)}
\label{appendix:vrf}
A standard VRF protocol comprises two entities: a prover and a verifier~\cite{papadopoulos2017making}. The prover generates a VRF output \( \beta = \text{HASH}(sk, \alpha) \) and an associated proof \( \pi = \text{Proof}(sk, \alpha) \), where \( sk \) denotes the secret key and \( \alpha \) is an input provided by the user. These values, \( \beta \) and \( \pi \), are sent to the verifier. The verifier initially checks whether \( \beta \) is consistent with the proof by evaluating \( \beta = \text{VRF\_proof\_to\_hash}(\pi) \). A mismatch at this stage suggests a potential security breach. To authenticate further, the prover reveals their public key \( pk \), enabling the verifier to perform a formal check using \( \text{Verify}(pk, \alpha, \pi) \). This function returns true if the proof \( \pi \) correctly validates \( \beta \) with respect to \( pk \) and \( \alpha \); otherwise, it returns false.

\section{Lookup Table for Minimum Client Selection Threshold in Cluster-oriented Setting}
\label{appendix:min-c-table}

\begin{table}[h]
\centering
\caption{Minimum required client selection threshold \( C \) to ensure \( P(\text{honest clients} < 2) \leq \delta \), for various values of collusion probability \( \phi \).}
\begin{tabular}{|c|c|c|c|c|c|}
\hline
\( \delta \backslash \phi \) & 0.1 & 0.2 & 0.3 & 0.4 & 0.5 \\
\hline
0.05  & 3  & 4  & 5  & 6  & 8  \\
0.01  & 4  & 5  & 7  & 8  & 11 \\
0.001 & 5  & 7  & 9  & 11 & 14 \\
\hline
\end{tabular}
\label{tab:min_c_values}
\end{table}

Each entry in Table~\ref{tab:min_c_values} represents the minimum required client selection threshold \( C \) for a given combination of collusion probability \( \phi \) and risk tolerance \( \delta \). Specifically, each cell gives the smallest value of \( C \) such that the probability of selecting fewer than 2 honest clients from a group of \( C \) selected clients is at most \( \delta \). This assumes that each client in the cluster independently colludes with the aggregator with probability \( \phi \). Practitioners can use this table to choose an appropriate value of \( C \) by identifying their system's estimated collusion probability and desired level of privacy assurance.

\section{Convergence Analysis}
\label{appendix:convergence}
To thoroughly analyze the convergence of AdRo-FL under static quantization (referring to section \ref{optimization}), we begin by introducing following assumptions:

\begin{assumption}[Bounded Gradients]
We assume that the local gradients for all clients are bounded, which means there exists a constant \(G\) such that:
\begin{equation}
\|\nabla f_i(\Theta)\|_2 \leq G, \quad \forall i, \forall \Theta.
\end{equation}
This assumption prevents excessively large updates, ensuring stability during the optimization process.
\end{assumption}

\begin{assumption}[Bounded Variance of Stochastic Gradients]
We assume the variance of gradients computed by the selected subset \(S_t\) of clients at round \(t\) is bounded as follows:
\begin{equation}
\mathbb{E} \left[ \left\| \frac{1}{|S_t|} \sum_{i \in S_t} \nabla f_i(\Theta) - \nabla F(\Theta) \right\|_2^2 \right] \leq \frac{\sigma^2}{|S_t|}.
\end{equation}
This assumption ensures the updates are not excessively noisy.
\end{assumption}

\begin{assumption}[\(L\)-Smoothness of the Global Loss]
We assume the global loss function \(F(\Theta)\) is \(L\)-smooth, which implies:
\begin{equation}
\|\nabla F(\Theta) - \nabla F(\Theta')\|_2 \leq L \|\Theta - \Theta'\|_2, \quad \forall \Theta, \Theta'.
\end{equation}
This guarantees smooth and gradual changes in the gradient, aiding stable convergence.
\end{assumption}

\begin{assumption}[Selection Bias]
We consider the bias \(R_t\) introduced by client selection at each round:
\begin{equation}
\mathbb{E}\left[\frac{1}{|S_t|} \sum_{i \in S_t}\nabla f_i(\Theta)\right] = \nabla F(\Theta) + R_t.
\end{equation}
Here, \(R_t\) represents the bias due to the non-uniform selection of clients.
\end{assumption}

\subsection{Step 1: Update Rule with Static Quantization}

The global model update incorporating static quantization is:
\begin{equation}
\Theta^{t+1} = \Theta^t - \eta \frac{1}{|S_t|} \sum_{i \in S_t} \nabla f_i(\Theta^t),
\end{equation}
where \(\eta\) is the learning rate. The static quantization affects only the transmitted payload size, not explicitly the gradients in this formulation.

\subsection{Step 2: Applying the \(L\)-Smoothness Property}

Applying the \(L\)-smoothness definition to our update step, we obtain:
\begin{align}
F(\Theta^{t+1}) &\leq F(\Theta^t) + \nabla F(\Theta^t)^T(\Theta^{t+1} - \Theta^t) + \frac{L}{2}\|\Theta^{t+1}-\Theta^t\|^2 \\
&= F(\Theta^t) - \eta \nabla F(\Theta^t)^T \frac{1}{|S_t|}\sum_{i \in S_t}\nabla f_i(\Theta^t) \\
&\quad + \frac{L \eta^2}{2}\left\|\frac{1}{|S_t|}\sum_{i \in S_t}\nabla f_i(\Theta^t)\right\|^2.
\end{align}
This describes how the loss changes at each iteration, highlighting the effect of the learning rate and gradients.

\subsection{Step 3: Taking Expectation}

Taking the expectation conditioned on the current parameters \(\Theta^t\), we have:
\begin{align}
\mathbb{E}[F(\Theta^{t+1})|\Theta^t] &\leq F(\Theta^t) - \eta \nabla F(\Theta^t)^T\left(\nabla F(\Theta^t)+R_t\right) \\
&\quad + \frac{L \eta^2}{2}\mathbb{E}\left[\left\|\frac{1}{|S_t|}\sum_{i \in S_t}\nabla f_i(\Theta^t)\right\|^2\right].
\end{align}
This accounts explicitly for bias due to client selection.

\subsection{Step 4: Bounding the Gradient Norm Term}

Using bounded variance and gradient assumptions, we bound the gradient norm as follows:
\begin{align}
\mathbb{E}\left[\left\|\frac{1}{|S_t|}\sum_{i \in S_t}\nabla f_i(\Theta^t)\right\|^2\right] &\leq \frac{\sigma^2}{|S_t|} + \|\nabla F(\Theta^t)\|^2.
\end{align}
This step ensures stable updates despite the stochastic nature of client sampling.

\subsection{Step 5: Deriving the Final Convergence Bound}

Combining the above results, the final expected decrease in the loss becomes:
\begin{align}
\mathbb{E}[F(\Theta^{t+1})|\Theta^t] &\leq F(\Theta^t) - \eta\|\nabla F(\Theta^t)\|^2 + \eta \|\nabla F(\Theta^t)\|\|R_t\| \\
&\quad + \frac{L \eta^2}{2}\left(\frac{\sigma^2}{|S_t|}+\|\nabla F(\Theta^t)\|^2\right).
\end{align}

This explicitly captures the effects of bias \(R_t\), variance \(\sigma^2\), and learning rate \(\eta\) on the convergence. 

Provided that the selection bias \(R_t\) diminishes or remains small, and given appropriate choices of \(\eta\), the global model parameters \(\Theta\) will converge towards a stationary point. 

%icon credit: https://www.flaticon.com/

\end{document}